# The PomXYZ proteins self-organize on the bacterial nucleoid to stimulate cell division

Dominik Schumacher[1], Silke Bergeler[2], Andrea Harms[1], Janet Vonck[3], Sabrina Huneke[1], Erwin Frey[2] & Lotte Søgaard-Andersen[1, 4]

[1] Department of Ecophysiology, Max Planck Institute for Terrestrial Microbiology, Karl-von-Frisch Str. 10, 35043 Marburg, Germany

[2] Arnold Sommerfeld Center for Theoretical Physics and Center for NanoScience, Department of Physics, Ludwig-Maximilians-Universität München, Theresienstr. 37, 80333 Munich, Germany.

[3] Department of Structural Biology, Max Planck Institute of Biophysics, 60438 Frankfurt am Main, Germany.

[4] Corresponding author
Tel. +49-6421-178 201
Fax +49-6421-178 209
E-mail: sogaard@mpi-marburg.mpg.de

**Summary**

Cell division site positioning is precisely regulated to generate correctly sized and shaped daughters. We uncover a novel strategy to position the FtsZ cytokinetic ring at midcell in the social bacterium *Myxococcus xanthus.* PomX, PomY and the nucleoid-binding ParA/MinD ATPase PomZ self-assemble forming a large nucleoid-associated complex that localizes at the division site before FtsZ to directly guide and stimulate division. PomXYZ localization is generated through self-organized biased random motion on the nucleoid towards midcell and constrained motion at midcell. Experiments and theory show that PomXYZ motion is produced by diffusive PomZ fluxes on the nucleoid into the complex. Flux differences scale with the intracellular asymmetry of the complex and are converted into a local PomZ concentration gradient across the complex with translocation towards the higher PomZ concentration. At midcell, fluxes equalize resulting in constrained motion. Flux-based mechanisms may represent a general paradigm for positioning of macromolecular structures in bacteria.

**Introduction**

Spatiotemporal organization of cellular content is a fundamental property of all cells and central to their proper functioning. Regulatory mechanisms setting up this organization operate to target proteins to specific subcellular positions, thus, spatially confining their activities (Nelson, 2003; Shapiro et al., 2009; Treuner-Lange and Søgaard-Andersen, 2014). Positioning of the cell division site requires exquisite spatiotemporal control to ensure the generation of daughter cells of the correct size, shape and chromosome complement.

In bacteria, cell division initiates with the assembly of the highly conserved tubulin-like protein FtsZ into a ring-like structure, the cytokinetic Z-ring, at the future division site (Bi and Lutkenhaus, 1991). The Z-ring directly or indirectly recruits the remaining proteins of the cytokinetic machinery (Lutkenhaus et al., 2012). Consistently, systems that regulate positioning of the cell division site control Z-ring formation and position (Lutkenhaus et al., 2012). The core proteins of the cytokinetic machinery are highly conserved in different bacterial lineages. By contrast, the systems that regulate when and where the Z-ring forms are diverse and still incompletely understood. In the rod-shaped cells of *Escherichia coli*, *Bacillus subtilis* and *Caulobacter crescentus* that all divide at midcell, placement of the Z-ring relies on systems that inhibit Z-ring formation throughout cells except at midcell. In *E. coli* and *B. subtilis,* the Min system and nucleoid occlusion work in concert to regulate Z-ring formation. Nucleoid occlusion depends on DNA binding proteins that inhibit Z-ring formation over the nucleoid and couple Z-ring formation to completion of chromosome replication and segregation (Adams and Errington, 2009). Proteins of the Min system inhibit Z-ring formation at the cell poles. In *B. subtilis* these proteins are recruited to the cell poles by DivIVA and the adaptor MinJ (Treuner-Lange and Søgaard-Andersen, 2014). By contrast, proteins of the Min system in *E. coli* self-organize (Howard et al., 2001; Meinhardt and de Boer, 2001; Kruse, 2002; Huang et al., 2003; Fange and Elf, 2006; Touhami et al., 2006; Loose et al., 2008; Halatek and Frey, 2012) to undergo coupled pole-to-pole oscillations (Hu and Lutkenhaus, 1999; Raskin and de Boer, 1999). In this system, MinD, which is a member of the ParA/MinD superfamily of P-loop ATPases, in its ATP-bound dimeric form, cooperatively binds to the cytoplasmic membrane and forms a complex with MinC that inhibits Z-ring formation (Hu and Lutkenhaus, 1999; Hu et al., 1999; Hu and Lutkenhaus, 2001; Hu et al., 2002; Lackner et al., 2003). MinD also recruits its ATPase Activating Protein (AAP) MinE to the membrane triggering ATPase activity and membrane unbinding of MinD in its monomeric form. Subsequently, MinD undergoes nucleotide exchange and rebinds to the membrane (Hu and Lutkenhaus, 2001; Hu et al., 2002; Lackner et al., 2003). In the absence of MinE, MinD-ATP/MinC accumulates along the entire membrane blocking Z-ring assembly throughout cells (Raskin and de Boer, 1999). However, in the presence of MinE, coupled pole-to-pole oscillations of the MinD-ATP/MinC complex and MinE occur. Because the lowest concentration of MinD-ATP/MinC integrated over time is at midcell, these oscillations limit assembly of the Z-ring to midcell (Meinhardt and de Boer, 2001). In *C. crescentus*, the ParA/MinD ATPase MipZ directly inhibits FtsZ polymerization (Thanbichler and Shapiro, 2006). MipZ is recruited to the cell poles by ParB and forms bipolar gradients on the nucleoid extending from the poles towards midcell effectively restricting cell division to midcell (Kiekebusch et al., 2012). In contrast to these systems that all incorporate a ParA/MinD ATPase and inhibit Z-ring formation, the SsgB/SsgA proteins in *Streptomyces coelicolor* and MapZ in *Streptococcus pneumoniae* directly recruit FtsZ to the division site. SsgB/SsgA dictate the site of Z-ring formation during the multiple synchronous cell divisions in aerial mycelia (Willemse et al., 2011). How SsgB/SsgA identifies future division sites is not known. In the ovoid *S. pneumoniae* cells, the integral membrane protein MapZ localizes to midcell together with FtsZ. During cell growth at midcell, MapZ is moved passively from midcell to the future division sites in each of the two daughters (Fleurie et al., 2014; Holeckova et al., 2015).

ParA/MinD ATPases are highly versatile and not only involved in the spatiotemporal control of cell division but also in chromosome and plasmid segregation and positioning of other cellular structures including clusters of chemosensory proteins, proteinaceous microcompartments and flagella (Gerdes et al., 2010; Lutkenhaus, 2012). Among these systems, those involved in chromosome and plasmid segregation are best understood. Here, ParA dimerizes upon ATP binding and binds nonspecifically to the nucleoid while the monomeric ADP-bound and apo-forms are cytosolic (Leonard et al., 2005; Hester and Lutkenhaus, 2007; Scholefield et al., 2011). The low intrinsic ATPase activity of ParA is stimulated by the AAP ParB, a MinE analog (Easter and Gober, 2002; Leonard et al., 2005; Barillà et al., 2007; Ah-Seng et al., 2009; Scholefield et al., 2011). ParB binds to *parS* sequences close to the chromosomal origin of replication or on a plasmid (Livny et al., 2007). ParB/*parS* segregation occurs by a process in which this complex interacts with ParA dimers that are bound nonspecifically to the nucleoid. Because this interaction stimulates ParA ATPase activity it results in ParA dissociation from the nucleoid (Fogel and Waldor, 2006; Ptacin et al., 2010; Schofield et al., 2010; Vecchiarelli et al., 2013). Subsequently, the ParB/*parS* complex interacts with flanking nucleoid-bound ParA dimers. Repeated cycles of these events result in the translocation of the ParB/*parS* complex across the nucleoid and a zone depleted of ParA is generated in the wake of the translocating complex. The directionality of translocation is thought to be determined by this concentration gradient of nucleoid-bound ParA that spans across the entire nucleoid (Fogel and Waldor, 2006; Ringgaard et al., 2009; Ptacin et al., 2010; Schofield et al., 2010). In *C. crescentus* formation of the ParA gradient involves the sequestration of ParA monomers by the polar landmark proteins PopZ and TipN (Ptacin et al., 2010; Schofield et al., 2010). In the case of plasmids, slow nucleotide exchange and rebinding to the nucleoid by ParA are thought to contribute to formation of this gradient (Vecchiarelli et al., 2010).

Regulation of Z-ring formation and cell division at midcell in the rod-shaped cells of the social bacterium *Myxococcus xanthus* depend on the ParA/MinD P-loop ATPase PomZ, which lacks the C-terminal amphipathic helix that associates MinD to the membrane (Treuner-Lange et al., 2013). PomZ has a unique localization pattern among characterized ParA/MinD ATPases (Treuner-Lange et al., 2013): Upon cell division, PomZ forms a cluster over the nucleoid; later in the cell cycle, this cluster localizes at midcell and here PomZ colocalizes with FtsZ. Intriguingly, PomZ localizes to midcell before as well as in the absence of FtsZ suggesting that PomZ could be part of a spatiotemporal control system that directly recruits FtsZ to midcell. However, the molecular mechanisms mediating midcell localization of PomZ and stimulation of Z-ring formation have so far remained unknown.

Here, we identify the two previously uncharacterized proteins, PomX and PomY, as important for midcell localization of PomZ and show that they function in concert with PomZ to directly recruit FtsZ to midcell and stimulate Z-ring formation. We demonstrate that the PomXYZ proteins self-assemble to form a large complex on the nucleoid that translocates to the midnucleoid, which coincides with midcell, in a biased random walk that depends on nucleoid binding and PomXY-stimulated ATP-hydrolysis by PomZ. At the midnucleoid the complex undergoes constrained motion and stimulates Z-ring formation. By combining experimental work and theory, we provide evidence that the directionality and constrained motion of this complex arise from a mechanism that depends on the diffusive flux of nucleoid-bound PomZ into the PomXYZ cluster as previously suggested for equipositioning of plasmids and protein complexes (Ietswaart et al., 2014). Specifically, the diffusive PomZ fluxes into the cluster from either side along the long cell axis scale with the intracellular asymmetry of the cluster and convert this asymmetry into a local PomZ concentration gradient across the PomXYZ complex and translocation of the complex towards the higher PomZ concentration. At midnucleoid, the diffusive PomZ fluxes into the cluster equalize, resulting in constrained motion. These analyses support a novel mechanism for how bacteria identify the site of cell division at midcell and explain how interactions at the molecular scale are transformed into cellular organization at the µm scale.

**Results**

<u>PomX and PomY are important for cell division, Z-ring formation and positioning</u>
While searching for proteins important for midcell localization of PomZ, we noticed that *pomZ* (MXAN_0635) is flanked by conserved genes in myxobacterial genomes (Fig. S1A). The predicted gene products are rich in protein-protein interaction domains with MXAN_0634 (henceforth PomY for <u>Po</u>sitioning at <u>m</u>idcell of FtsZ <u>Y</u>) containing an N-terminal coiled-coil domain, HEAT domain repeats and a proline-rich region while MXAN_0636 (henceforth PomX for <u>Po</u>sitioning at <u>m</u>idcell of FtsZ <u>X</u>) consists of a long N-terminal region without annotated domains and a C-terminal coiled-coil region (Fig. S1B).

Mutants with in-frame deletions in Δ*pomX* or Δ*pomY* phenocopied the Δ*pomZ* mutant with a growth rate comparable to wild-type (WT), formation of long filamentous cells and short anucleate minicells, and fewer cell division constrictions that were distributed along the cell length but did not occur over the nucleoid (Fig. 1A-C; S1CD). All double and the triple mutants phenocopied the mutants with single mutations (Fig. 1A-C; S1C), suggesting that PomXYZ function together. The cell division defects in the ΔpomX and ΔpomY mutants were complemented by ectopic expression at native or above native levels of mCherry-PomX and PomY-mCherry, respectively (Fig. 1B, S1CE). Moreover, PomX, PomY and PomZ accumulated independently of each other (Fig. 1D).

Cell division defects can be the result of chromosome replication or segregation defects. The ΔpomX and ΔpomY mutants had the same number of nucleoids per cell length as untreated WT and WT treated with the cell division inhibitor cephalexin to generate filamentous cells. Also, using the midpoint of the nucleoid (henceforth, referred to as the midnucleoid) as a marker, these nucleoids localized similarly (Fig. S2AB). Using a ParB-YFP fusion (Harms et al., 2013) as a marker for the origin of replication, the number and localization of origins of replication were similar in the ΔpomX and ΔpomY mutants to those in untreated and cephalexin-treated WT (Fig. S2AC). We conclude that PomX and PomY, similarly to PomZ (Treuner-Lange et al., 2013), are not important for replication and chromosome segregation.

All three pom mutants accumulated FtsZ at WT levels (Fig. 1E). By contrast, FtsZ localization was abnormal in the absence of PomX and PomY. As previously shown (Treuner-Lange et al., 2013), in 50% of WT cells FtsZ-mCherry was diffusely localized in the cytoplasm and in the remaining cells it formed a single cluster at midcell (defined as 50±5% of cell length) that spanned the cell width and corresponds to the Z-ring (Fig. 1F). In ΔpomX and ΔpomY cells, FtsZ-mCherry predominantly localized in the diffuse pattern and only 2-5% of cells had a cluster spanning the cell width and these clusters were not restricted to midcell but localized somewhere along the cell length (Fig. 1F). These observations are similar to those made for the ΔpomZ mutant (Treuner-Lange et al., 2013) suggesting that PomX, PomY and PomZ function together to stimulate Z-ring formation and positioning at midcell.

PomX, PomY and PomZ form a complex that is positioned at midcell by PomZ

To uncover the function of PomX and PomY in Z-ring formation and cell division, we determined their subcellular localization using active mCherry-PomX and PomY-mCherry fusions expressed at native levels. Overall, the two proteins showed the same localization pattern as PomZ [(Treuner-Lange et al., 2013); Fig. 2A]. mCherry-PomX and PomY-mCherry gave no signal or a diffuse signal in 10-26% of cells, formed a single cluster in an off-centre position (defined as clusters outside of the midcell region at 50±5% of cell length) in 22-34% of cells, and a cluster at midcell in the remaining 52-56% of cells. In all cells, off-centre clusters colocalized with the nucleoid (Fig. 2A). In the case of midcell clusters, ~75% localized over the midnucleoid (Fig. 2A, third row) and the remaining ~25% localized between two fully segregated nucleoids (Fig. 2A, fourth row). Moreover, PomX and PomY colocalized with cell division constrictions (Fig. S3A).

In addition to forming a cluster, a significant fraction (~90%) of PomZ in all cells colocalized with the nucleoid generating a patchy localization pattern (Fig. S3B) suggesting that PomZ binds nonspecifically to the nucleoid. The patchy PomZ signal over the nucleoid is almost symmetrically distributed around the cluster as indicated by an asymmetry measure normalized for nucleoid area of 0.09±0.04 (n=19) for off-centre clusters and 0.05±0.04 (n=44) for midcell clusters (Supplemental Experimental Procedures; Fig. S3B). This asymmetry is slightly but significantly higher in the case of cells with an off-centre cluster (P < 0.05, t-test) and with the highest intensity on the side of the cluster containing most of the nucleoid. For comparison, the same asymmetry measure for Pico Green stained nucleoids is 0.07±0.06 (n=49).

PomY-YFP/PomZ-mCherry and PomY-YFP/mCherry-PomX perfectly colocalized in the off-centre clusters and at the midnucleoid at midcell (Fig. 2B). Using an Ssb-YFP fusion as a proxy for assembled replisomes (Harms et al., 2013), we also observed that PomX and PomY localized over the midnucleoid at midcell while replication was ongoing (Fig. S3C). Altogether, these findings demonstrate that PomXYZ early in the cell cycle colocalize in an off-centre position on the nucleoid, later at the midnucleoid at midcell, and this midcell localization persists at least until cell division initiates.

We hypothesized that if PomXYZ interact to form a complex, then lack of one of the proteins would perturb complex formation and/or localization. To this end, we systematically localized each Pom protein in the absence of one or the other Pom protein (Fig. 2C). mCherry-PomX formed clusters and localized independently of PomY; however, the clusters had an aspect ratio of 3.6±2.9 compared to 1.2±0.2 in the presence of PomY. mCherry-PomX also formed clusters independently of PomZ; however, these clusters were rarely at midcell and 37% of clusters were in the large nucleoid-free subpolar regions (Fig. S3D). By contrast, PomY-mCherry was dispersed in the absence of PomX and formed slightly fewer clusters in the absence of PomZ and these clusters were rarely at midcell and frequently in the nucleoid-free subpolar regions (Fig. S3D). PomZ-mCherry did not form clusters and localized in the patchy pattern over the nucleoid in the absence of PomX. In the absence of PomY, PomZ-mCherry also mostly localized in the patchy pattern over the nucleoid and only formed a few clusters and ~35% of these were at midcell (Fig. 2C). Finally, mCherry-PomX and PomY-YFP colocalized in the absence of PomZ (Fig. S3E).

Altogether, these observations are consistent with PomX nucleating the formation of a complex that contains all three Pom proteins and with PomZ being central to localization of this complex at midcell (Fig. 2D). From the frequent localization of the PomXY complex to the nucleoid-free subpolar regions in the absence of PomZ, we infer that PomZ also associates this complex with the nucleoid.

PomX and PomY localize to midcell before FtsZ
To address the causal relationship between midcell localization of PomX/PomY and FtsZ, we localized FtsZ-GFP expressed at native levels in the presence of unlabeled FtsZ (Treuner-Lange et al., 2013) in strains that also expressed mCherry-PomX or PomY-mCherry. We observed several patterns of PomX/FtsZ and PomY/FtsZ localization. Importantly, in a large fraction of cells, mCherry-PomX (49%) or PomY-mCherry (28%) were at midcell without FtsZ-GFP and we did not observe the opposite pattern (Fig. 3A). FtsZ-GFP was perfectly superimposable with mCherry-PomX and PomY-mCherry at midcell but not in the off-centre position (Fig. 3A). These observations are in agreement with previous findings that PomZ localizes to midcell before FtsZ and once both proteins are at midcell they perfectly colocalize (Treuner-Lange et al., 2013).

We examined the localization of mCherry-PomX and PomY-mCherry in cells depleted for FtsZ, by expressing the only copy of *ftsZ* from a $Cu^{2+}$ inducible promoter at an ectopic site. In the presence of $Cu^{2+}$ the two strains displayed normal cell length distributions and constriction frequencies (Fig. 3B). As described (Treuner-Lange et al., 2013), in the absence of $Cu^{2+}$, FtsZ accumulation decreased over time and was not detectable in immunoblots after 6-9 hrs. In parallel, the frequency of cell division constrictions decreased and cell length increased. After 12 hrs of FtsZ depletion, all cells contained mCherry-PomX or PomY-mCherry clusters (Fig. 3B; S4A). Importantly, these clusters often localized at midcell. Thus, similarly to PomZ (Treuner-Lange et al., 2013), PomX and PomY localize at midcell in the absence of FtsZ. In control experiments, we observed that in cells treated with cephalexin, PomX and PomY also remained at midcell in a large fraction of cells (Fig. S4B). As expected, we also observed that the cell division protein FtsK, which is recruited late to the cytokinetic machinery in an FtsZ-dependent manner (Lutkenhaus et al., 2012), did not form midcell clusters after depletion of FtsZ (Fig. S4C). Altogether, these data suggest that the PomXYZ complex localizes at midcell independently of FtsZ and they are consistent with a model in which these three proteins function together to recruit FtsZ to midcell.

PomY and PomZ interact directly with FtsZ

To explore how PomXYZ stimulate Z-ring formation at midcell, we determined whether any of them interact directly with FtsZ. To this end, we carried out a yeast two hybrid screen for binary interactions. In these analyses, FtsZ self-interacted; moreover, PomY as well as PomZ interacted with FtsZ (Fig. 3C). To confirm that the self-interaction observed for FtsZ results in polymerization in a GTP-dependent manner as has been shown for many FtsZ proteins, we overexpressed and purified soluble native FtsZ (Fig. S4D). Purified FtsZ formed higher order structures in the presence of GTP as shown by right angle light scattering (Fig. S4E). FtsZ also formed filaments as shown by negative stain transmission electron microscopy (EM) in a GTP-dependent manner (Fig. S4F) similarly to other FtsZ proteins analyzed *in vitro*. Of note, we previously reported that FtsZ from *M. xanthus* has cooperative GTPase activity *in vitro* but did not form filaments visible by right angle light scattering and EM (Treuner-Lange et al., 2013). Here, we used a slightly different purification procedure and performed the polymerization assays at a slightly lower pH than in our previous analyses. We attribute the different results to these differences in experimental setups.

To test *in vivo* for the relevance of the interactions between PomY or PomZ and FtsZ, we asked whether the few cell divisions in Δ*pomY* cells and Δ*pomZ* cells occur over the mCherry-PomX cluster. Only 20% of the few cell divisions in the Δ*pomY* mutant occurred over mCherry-PomX clusters some of which also contain PomZ (Fig. 3D; Cf. Fig. 2C) while all cell divisions in the Δ*pomZ* mutant occurred over the mCherry-PomX cluster all of which contain PomY (Fig. 3D; Cf. Fig. 2C; S3E). We conclude that FtsZ interacts directly with PomY and PomZ. Moreover, our data strongly suggest that PomY in the PomXYZ complex is essential for recruiting FtsZ to the site of cell division and that all three Pom proteins are important for efficient Z-ring formation.

The PomXYZ complex relocates to midcell by PomZ-dependent translocation

We performed time-lapse microscopy to resolve how the PomXYZ complex shifts from an off-centre to a midnucleoid position at midcell. In time-lapse recordings with images recorded every 15 min we observed the same pattern for PomX and PomY (Fig. 4A). Starting with a cluster at midcell, this cluster splits into two during cell division resulting in two daughter cells with off-centre clusters close to the new cell pole. Subsequently, each cluster slowly migrated to midcell in a daughter, and then remained there. As expected, PomX and PomY colocalized during translocation (Fig. S5A) and the translocation time to midcell after release from a division site was similar for the two proteins (Fig. 4A). Occasionally, PomX and PomY were asymmetrically distributed to the daughters (Fig. S5B). We speculate that this asymmetric distribution gives rise to the cells with no or a diffuse signal of the Pom proteins in snapshots (Cf. Fig. 2A). The generation time of *M. xanthus* cells under the conditions of this experiment is ~5 hrs, replication initiates immediately after cell division and takes ~4 hrs (Harms et al., 2013). Thus, the PomXYZ complex localizes at midnucleoid at midcell, on average, 2-3 hrs before replication is complete. As expected, constrictions occurred over the mCherry-PomX and PomY-mCherry cluster in all cases observed (n=65). We conclude that PomXYZ complex localization to midcell depends on translocation from an off-centre position.

To gain further insights into this translocation mechanism, we monitored the PomY-mCherry cluster as a marker for the PomXYZ complex at higher temporal resolution (images every 30 sec for 20 min). To quantify cluster movement, we tracked the centroid of the cluster and generated two-dimensional trajectories for individual clusters. At this temporal resolution, most PomY-mCherry clusters moved along the long and the short axes of cells and did not follow a straight path (Fig. 4B). Qualitatively, cluster dynamics varied depending on whether a cluster was in an off-centre position or at midcell. Off-centre clusters displayed long periods of wandering towards midcell reminiscent of a two-dimensional biased random walk. By contrast, clusters in the midcell region while still moving had less directional bias. Finally, ~10% of midcell clusters, and these were mostly in long cells, essentially displayed no motion. We speculate that these cells are undergoing division and that the PomXYZ proteins are associated with the cytokinetic machinery and, therefore, display less motion.

To systematically quantify cluster motion, we calculated the mean cumulative squared distance (MCSD; sum of mean squared distances moved per 30 sec interval) and the mean squared displacement (MSD) from the PomY-mCherry cluster trajectories (Fig. 4C). Off-centre and midcell clusters moved similar MCSDs ($0.82\pm0.17\mu m^2$ and $0.59\pm0.13\mu m^2$ after 20 min, respectively). However, the MSD showed clear differences with the MSD for off-centre clusters ($0.47\pm0.70\mu m^2$ after 20 min) displaying a slope over time, which seems to increase, indicating that they exhibited directed motion whereas the MSD for midcell clusters ($0.11\pm0.16\ \mu m^2$ after 20 min) reached a plateau demonstrating that cluster motion was constrained to the midcell region.

Strikingly, lack of PomZ strongly reduced the MCSD of PomY-mCherry clusters ($0.19\pm0.04\mu m^2$ after 20 min) and the MSD ($0.07\pm0.20\mu m^2$ after 20 min) reached a plateau slightly lower than in the case of midcell clusters in a *pomZ*$^+$ background (Fig. 4BC). We conclude that PomZ is essential for cluster motion with translocation to midcell by a biased random walk and constrained motion at midcell. Also, in the absence of PomZ, the cluster is largely stalled somewhere in a cell. These observations also provide evidence that the mechanism underlying PomZ-dependent translocation of the PomXYZ complex is able to "sense" cluster position within cellular space and adjusts cluster motion accordingly.

The *M. xanthus* chromosome is arranged about a longitudinal axis with the origin of replication and the terminus region in the subpolar regions close to the old and new poles, respectively and with large (~1.5µm long) nucleoid-free subpolar regions [Cf. DAPI stained cells in Fig. 2AB, S2B; (Harms et al., 2013)]. During replication, one of the origins translocates to the subpolar region close to the new cell pole followed by the rest of the chromosome, and, in parallel, the terminus region displaces towards midcell. The movement of the terminus region is somewhat comparable to that of the PomXYZ cluster towards midcell. In principle, translocation of the PomXYZ complex to midcell could depend on specific binding to the terminus region in a process in which the PomXYZ complex would "piggyback" on the terminus. To this end, we quantified the dynamics of the terminus region using FROS (Fluorescence Repressor Operator System) with TetR-YFP bound to a *tetO* array at 192° on the *M. xanthus* chromosome (Harms et al., 2013). Similarly to the PomXY cluster in the absence of PomZ, this locus displayed very little motion (MCSD of 0.21±0.07µm$^2$ and MSD of 0.05±0.18µm$^2$ after 20 min) (Fig. 4BC), strongly suggesting that the PomXYZ complex is not "piggybacking" on the terminus to midcell.

PomX and PomY form a complex that stimulates ATPase activity by DNA bound PomZ

To determine how PomZ promotes the motion of the PomXYZ complex, we analyzed whether PomX, PomY and PomZ interact directly. Using the yeast two hybrid system, we observed that all three proteins self-interacted and interacted in all pairwise combinations (Fig. 3C). Next, we expressed the active Pom-fusion proteins alone or together in *E. coli* that lacks close relatives of the Pom proteins (Fig. 5A). In *E. coli*, PomZ-mCherry alone perfectly colocalized with the nucleoid without forming clusters, supporting the notion that PomZ binds nonspecifically to DNA. PomY-YFP displayed a diffuse signal throughout cells and often formed a polar cluster in nucleoid-free areas. mCherry-PomX formed small patches and longer filamentous structures in nucleoid-free areas. Co-expressed PomY-YFP and mCherry-PomX colocalized in filamentous patches whereas co-expressed PomZ-mCherry and PomY-YFP colocalized on the nucleoid without forming clusters. Because PomX/PomY and PomY/PomZ interacted and PomZ-mCherry is able to recruit PomY-YFP to the nucleoid in *E. coli*, we reasoned that co-expression of PomY-YFP and PomZ-mCherry together with unlabeled PomX would lead to reconstitution of the PomXYZ cluster on the *E. coli* nucleoid. Remarkably, in 64% of cells PomY-YFP colocalized with PomZ-mCherry on the nucleoids forming clusters with the same dimensions as those in *M. xanthus*. While *M. xanthus* cells contain a single PomXYZ cluster until it splits late in cell division, *E. coli* cells generally contained a cluster over each nucleoid. We speculate that more than one cluster is formed in *E. coli* because the Pom proteins are not associated with the cytokinetic machinery. We conclude that the three Pom proteins interact directly in all pairwise combinations. Moreover, all three proteins are required and sufficient for the formation of nucleoid-associated clusters suggesting that the PomXYZ proteins constitute a nucleoid-associated machine that is able to translocate over the nucleoid.

To confirm the interactions between the Pom proteins, we overexpressed and purified soluble full length PomX-His$_6$, PomY-His$_6$, and His$_6$-PomZ (Fig. S4D). After high-speed centrifugation, 90% of PomX-His$_6$ was recovered in the pellet fraction whereas PomY-His$_6$ was equally distributed in the pellet and soluble fractions (Fig. S6A). By contrast, PomY-His$_6$ mixed with an equimolar amount of PomX-His$_6$ was almost entirely recovered in the pellet fraction. In EM analyses, PomX-His$_6$ alone formed long thin filaments 8.3±1.9nm in width and a length of several µm (Fig. 5B) whereas PomY-His$_6$ under the same conditions did not form higher order structures. However, when mixed in a 1:1 molar ratio, PomX-His$_6$ and PomY-His$_6$ formed thick bundles up to 150nm in width and a length of several µm. Thin filaments emerged from these bundles suggesting that these structures consist of PomX-His$_6$ filaments bundled by PomY-His$_6$.

A PomZ$^{D90A}$ variant, which is predicted to be blocked in ATP hydrolysis, is non-functional *in vivo* [(Treuner-Lange et al., 2013); see below]. Therefore, we tested if PomX and/or PomY affect PomZ ATPase activity *in vitro* (Fig. 5C). As reported (Treuner-Lange et al., 2013), ATPase activity by His$_6$-PomZ alone was undetectable. Equimolar amounts of PomX-His$_6$ or PomY-His$_6$ added to His$_6$-PomZ did not detectably stimulate ATPase activity whereas PomX and PomY together stimulated ATP hydrolysis by PomZ resulting in a turnover rate of 18 ATP h$^{-1}$. Because PomZ-mCherry binds nonspecifically to the nucleoid in *M. xanthus* and in *E. coli*, contains the conserved amino acid residue important for nonspecific DNA binding by other ParA ATPases (Fig. S6H), and ATP hydrolysis by ParA ATPases involved in plasmid or chromosome segregation is stimulated by nonspecific DNA binding (Ah-Seng et al., 2009; Scholefield et al., 2011), we tested ATP hydrolysis by His$_6$-PomZ in the presence of nonspecific DNA. Under these conditions, ATPase activity by His$_6$-PomZ alone was weakly stimulated (1.6 ATP h$^{-1}$). PomX-His$_6$ and PomY-His$_6$ independently stimulated His$_6$-PomZ ATPase activity to 12 ATP h$^{-1}$; however, the two proteins synergistically stimulated PomZ-His$_6$ ATPase activity to 65 ATP h$^{-1}$. PomX and PomY do not contain DNA binding domains and do not appear to bind to the nucleoid *in vivo*. Therefore, these observations strongly suggest that PomX and PomY in the PomXY complex function synergistically to stimulate ATP hydrolysis by PomZ bound nonspecifically to DNA.

ATP-bound dimeric PomZ recruits the PomXY complex to the nucleoid
The ATPase cycle of ParA ATPases can be blocked at specific steps by specific substitutions (Leonard et al., 2005; Hester and Lutkenhaus, 2007; Ptacin et al., 2010; Schofield et al., 2010; Kiekebusch et al., 2012) (Fig. S6B). These substitutions in PomZ correspond to PomZ$^{K66Q}$ and PomZ$^{G62V}$, which are predicted monomeric variants, PomZ$^{D90A}$, which is predicted to be locked in the ATP-bound dimeric form that binds DNA nonspecifically, and PomZ$^{K268E}$, which is predicted to be blocked in DNA binding (Fig. S6H). To confirm that these PomZ variants have the predicted properties, we turned to the *E. coli* system (Fig. S6C). PomZ$^{D90A}$-mCherry, similarly to PomZ$^{WT}$-mCherry, perfectly colocalized with the nucleoid; PomZ-mCherry variants carrying the K66Q, G62V or K268E substitutions displayed diffuse localization, filling the nucleoid-free areas with fluorescent signal. We conclude that the ATP-bound dimeric PomZ binds the nucleoid nonspecifically whereas monomeric PomZ does not. The observation that PomZ$^{WT}$ colocalizes with the *E. coli* nucleoid also strongly suggests that PomZ$^{WT}$ spontaneously binds ATP and dimerizes.

Having confirmed that the PomZ variants have the predicted properties, we expressed them at native levels (Fig. S6E) and tested them for functionality in *M. xanthus*. Similar to the PomZ$^{D90A}$ fusion, neither PomZ$^{K66Q}$, PomZ$^{G62V}$ nor PomZ$^{K268E}$ fused to mCherry complemented the cell division defect of the Δ*pomZ* mutant (Fig. 5D). Notably, all three variants unable to bind DNA (PomZ$^{K66Q}$, PomZ$^{G62V}$, PomZ$^{K268E}$) failed to form clusters and only displayed diffuse localization in *M. xanthus* (Fig. S6D), demonstrating that they are unable to interact with PomXY to form a cluster. As described (Treuner-Lange et al., 2013), the ATP-locked, dimeric PomZ$^{D90A}$-mCherry variant formed a single cluster somewhere on the nucleoid (Fig. S6D). Most of PomZ$^{D90A}$-mCherry localized to this cluster with no or only little PomZ$^{D90A}$-mCherry binding nonspecifically to the nucleoid away from the cluster. The PomZ$^{D90A}$-mCherry cluster colocalized with PomY-YFP demonstrating that it is associated with the PomXY complex (Fig. S6F). Consistently, PomZ$^{D90A}$-mCherry cluster formation was absolutely dependent on PomX (Fig. S6G). Because PomZ$^{D90A}$-mCherry associates with the PomXY cluster but this cluster rarely localized at midcell, we characterized PomXYZ cluster dynamics in the presence of PomZ$^{D90A}$. As shown in Fig. 4BC, these clusters showed little motion (MCSD: 0.19±0.13μm$^2$, MSD: 0.05±0.06μm$^2$ after 20 min) similar to those in the Δ*pomZ* mutant.

Altogether, these data demonstrate that the conformation of PomZ that binds to the nucleoid, recruits the PomXY cluster to the nucleoid, and interacts with PomXY to generate the PomZ cluster, is the dimeric ATP-bound form. Moreover, ATP hydrolysis by PomZ is essential for PomXYZ cluster motion.

<u>PomZ is rapidly turned over in the PomXYZ cluster and highly dynamic on the nucleoid</u>
ATP-bound dimeric PomZ bound to the nucleoid interacts with the PomXY complex *in vivo*. However, PomXY stimulates ATP hydrolysis by PomZ when PomZ binds to DNA *in vitro*. If this stimulation also occurs *in vivo*, then the prediction is that PomZ rapidly turns over in the PomXYZ cluster. To test this prediction, we performed fluorescence recovery after photobleaching (FRAP) experiments. After short 60 msec laser pulses were applied to bleach PomZ-mCherry clusters, their fluorescence was restored within 9 sec with a half maximal recovery ($t_{1/2}$) of 1.2±0.2 sec (n=20) (Fig. 6A). During this 60 msec laser pulse, the PomZ-mCherry signal on the nucleoid outside of the cluster was also bleached (Fig. 6A) suggesting that PomZ is not only rapidly turned over in the cluster but also highly dynamic on the nucleoid. To this end, we bleached PomZ-mCherry in a small region on the nucleoid outside of a cluster for 60 msec. During this short laser pulse, the entire PomZ-mCherry signal on the bleached side of the nucleoid relative to the PomZ-mCherry cluster was reduced (Fig. 6A). Therefore, to quantify fluorescence recovery, the PomZ-mCherry signal was divided into three parts, i.e. the signal on the bleached side of the nucleoid relative to the cluster, the unbleached side of the nucleoid relative to the cluster, and the cluster (Fig. 6A). This analysis demonstrated that concurrent with the fast recovery of fluorescence signal on the bleached side of the nucleoid ($t_{1/2}$=1.7±0.4 sec; n=18), the signals in the cluster as well as on the unbleached side of the nucleoid lost intensity restoring a pre-bleach situation within 9 sec (Fig. 6A). Thus, unbleached PomZ-mCherry in the cluster and on the unbleached side rapidly exchange with proteins on the bleached side.

To analyze if ATP-bound PomZ dimers undergo diffusion on the nucleoid, we bleached a small region on the nucleoid in strains overexpressing PomZ$^{WT}$-mCherry or PomZ$^{D90A}$-mCherry >50-fold (Fig. S1G; Fig. 6B). PomZ$^{WT}$-mCherry and PomZ$^{D90A}$-mCherry in these two strains bind the nucleoid creating an intense signal colocalizing with the nucleoid and, therefore, the cluster is not visible (Fig. 6B). A 160 msec laser pulse resulted in bleaching of the signal on the nucleoid outside of the bleached region. The signal in the bleached region recovered within 30 sec with a half-maximal recovery of 5.3±0.3 sec (PomZ$^{WT}$) and 8.3±0.4 sec (PomZ$^{D90A}$). Because PomZ$^{D90A}$-mCherry is locked in the ATP-bound dimeric form that binds to the nucleoid, this signal recovery is the result of diffusion of the protein on the nucleoid and not the result of rebinding to the nucleoid of PomZ dimers generated from a pool of monomers after ATP hydrolysis. As PomZ$^{WT}$-mCherry and PomZ$^{D90A}$-mCherry show the same overall recovery kinetics these data strongly imply that dimeric ATP-bound PomZ$^{WT}$ diffuses rapidly on the nucleoid generating a diffusive PomZ flux on the nucleoid and that this flux contributes to signal recovery in the PomZ cluster and on the nucleoid.

To determine if the fast turnover of PomZ in the cluster depends on ATP hydrolysis, we analyzed cells expressing PomZ$^{D90A}$-mCherry at native levels. Because most of PomZ$^{D90A}$-mCherry is in the PomXYZ cluster, we adopted a bleaching approach in which a laser pulse was applied for 3 sec to the nucleoid outside of the cluster. After this pulse, the PomZ$^{D90A}$-mCherry signal was unaffected (Fig. 6C). By contrast, the total cellular PomZ$^{WT}$-mCherry signal was strongly decreased after the 3 sec laser pulse demonstrating that most PomZ$^{WT}$ molecules had passed through the bleached area within 3 sec and confirming that PomZ$^{WT}$-mCherry is highly dynamic. As expected, the signals from the diffusely localized PomZ$^{K268E}$-mCherry, PomZ$^{K66Q}$-mCherry and PomZ$^{G62V}$-mCherry variants were also almost completely bleached during the 3 sec bleach (Fig. 6C). We conclude that ATP-locked dimeric PomZ is stably bound in the PomXYZ cluster.

A computational model for PomZ-dependent translocation and positioning of the PomXYZ complex

To understand the emergent properties of Pom system, i.e. how the local protein/protein/DNA interactions in the PomXYZ system are converted into a global cellular positioning system, we searched for experimentally-based mechanisms that would give rise to a biased random walk of the PomXYZ complex to midnucleoid and constrained motion at midnucleoid at midcell (Fig. 7A).

One model for ParA-dependent ParB/*parS* translocation is based on a diffusion-ratchet mechanism (Vecchiarelli et al., 2013; Vecchiarelli et al., 2014). A crucial aspect of this model is that the ParB/*parS* complex motion is slowed down by the interaction with DNA-bound ParA. However, we observe that the PomXY complex in the absence of PomZ is essentially non-motile and that PomZ drives the motion of the complex (Fig. 4BC). Therefore, we did not consider this model further. A second model for ParA-dependent ParB/*parS* translocation is based on the notion that ParA forms nucleoid associated filaments (Ringgaard et al., 2009; Ietswaart et al., 2014); however, we have no evidence supporting filament formation by PomZ (Treuner-Lange et al., 2013). Recently, Lim *et al.* (Lim et al., 2014) proposed a DNA-relay model for chromosome segregation by ParAB*S* systems in which the dynamics of the chromosome is included by assuming that the DNA-bound ParA is not fixed, but wiggles around due to the elastic properties of DNA. Because ParA is highly asymmetrically localized forming a steep gradient across the entire nucleoid, this effectively leads to a net force exerted on the ParB/*parS* complex towards a higher ParA concentration and the ParB/*parS* complex translocating up the gradient of nucleoid-bound ParA. We observe that PomZ is almost symmetrically distributed on the nucleoid around the PomXYZ cluster. Therefore, we did also not consider this model further. Instead, to explain the intracellular patterning of the Pom proteins, we propose a PomZ flux-based mechanism that builds on the mechanism recently proposed for equal positioning of plasmids over the nucleoid (Ietswaart et al., 2014). As in the DNA-relay model, we propose that the forces to move the PomXYZ cluster are generated by the elastic properties of the involved biological structures.

In this model, we assume that the PomXY cluster is an object with a fixed composition of PomX and PomY molecules and with a molecular weight of ~15 MDa (Supplemental Information). Hence, we reduce the system to a model consisting of the PomXY cluster, the nucleoid, and PomZ. For simplicity, the nucleoid and the PomXY cluster are modeled as one-dimensional lattices on which PomZ dimers can diffuse. In this model (Fig. 7B, S7A), ATP-bound PomZ dimers use the nucleoid as a scaffold to which they attach (1) and rapidly diffuse on (2). When bound to the nucleoid, PomZ dimers can also bind to and interact with the PomXY cluster (3). We hypothesize that PomZ dimers that are doubly bound to the nucleoid and PomXY can diffuse relative to both (4). ATPase activity of PomZ is stimulated when in contact with both the nucleoid and PomXY and after ATP hydrolysis, ADP-bound monomers are released (5) and undergo fast diffusion in the cytosol (6). Importantly, before PomZ can reattach to the nucleoid, the monomers have to undergo nucleotide exchange and dimerize giving rise to a time delay (clock symbol). Finally, the PomXY cluster is a barrier for PomZ dimers diffusing on the nucleoid. This might be due to the cluster acting as a PomZ sink and possibly as a steric hindrance. Because PomZ dimers diffuse rapidly on the nucleoid (and much faster than the cluster moves) and the PomXY cluster acts as a sink for diffusing PomZ dimers on the nucleoid, the difference in the PomZ flux into the cluster from the two sides along the long cell axis depends on the position of the cluster (Ietswaart et al., 2014). In the case of an off-centre cluster (Fig. 7B), more PomZ dimers arrive at the cluster from the side with the longer distance to the nucleoid end (i.e., from the right in Fig. 7B). This flux difference translates into an asymmetric concentration profile of PomZ dimers bound to the cluster, i.e. a local PomZ concentration gradient across the cluster, with the highest concentration on the side facing most of the nucleoid (Fig. 7CD). In the case of a cluster at midnucleoid, the PomZ fluxes into the cluster from the two sides equalize and the concentration profile of PomZ over the cluster is symmetric (Fig. 7CD).

It is not known how PomZ dimers generate the force to move the cluster. However, as recently pointed out, the chromosome has elastic properties (Wiggins et al., 2010; Lim et al., 2014) that can be harnessed to relay a ParB/*parS* complex across the steep gradient of nucleoid-bound ParA dimers (Lim et al., 2014). There is also evidence that proteins can act as elastic force bearing structures (Dietz and Rief, 2008). In our model, we effectively account for both sources of elasticity by modelling PomZ dimers as springs. Therefore, this aspect of our model is similar to the DNA-relay mechanism (Lim et al., 2014). In this scenario, there is a tug of war between PomZ dimers arriving from the left and right into the cluster: Some of the PomZ dimers arriving at the cluster from the right will bind to the cluster in a stretched conformation and generate a mechanical force that points to the right.

Similarly, PomZ dimers arriving from the left will generate a force pointing to the left. Therefore, any flux imbalance leads to a net force in the direction of the higher PomZ flux. In the case of off-centre clusters, this force points towards midnucleoid. In the case of clusters at midnucleoid, the PomZ flux from both sides is equal, and, therefore, no net force is exerted on the cluster (Fig. 7CD). Importantly, as long as the PomXY cluster moves in the direction of the highest PomZ concentration at the cluster, the proposed mechanism will result in biased random motion towards midnucleoid for off-centre clusters and constrained motion for clusters at midnucleoid, independently of the precise molecular mechanism giving rise to the forces resulting in cluster motion. Remarkably, this model reproduces off-centre to midnucleoid relocalization as well as constrained motion at midnucleoid of the PomXYZ cluster for physiological relevant parameters (Table S1, S2) with a timing similar to that observed *in vivo* (Fig. 7C and S7BC; Cf. Fig. 4AB).

The time delay between detachment of PomZ monomers from the cluster and subsequent reattachment of PomZ dimers to the nucleoid is important to guarantee that a PomZ flux imbalance into the cluster correlates with the asymmetry of the cluster on the nucleoid. If PomZ monomers would regain the ability to bind DNA quickly, they would bind to the nucleoid in close proximity to the cluster resulting in a decrease in the flux difference of PomZ dimers into the cluster from the two sides. In the *E. coli* Min system, a sufficiently small nucleotide exchange rate is also important for MinD channeling between the cell poles and, thus, robust pole-to-pole oscillations (Huang et al., 2003; Halatek and Frey, 2012).

Validating the model experimentally

The model predicts that PomZ overexpression would cause the mechanism to break down and reduce cluster motion (Supplemental Information; Fig. S7D). Indeed >50-fold overexpression of PomZ-mCherry causes a cell division phenotype similar to that of the Δ*pomZ* mutant (Treuner-Lange et al., 2013). Importantly, PomXYZ cluster motion in this mutant was significantly reduced (MCSD: 0.16±0.03μm$^2$, MSD: 0.09±0.18μm$^2$ after 20 min) similar to that in the Δ*pomZ* mutant (Fig. 4BC). The model further predicts that reduced PomZ ATPase activity would reduce the bias of the cluster motion towards midcell (Supplemental Information; Fig. S7E). Indeed, in the presence of PomZ$^{D90A}$ *in vivo*, cluster motion is strongly reduced (Fig. 4BC), in agreement with the model predictions. In the absence of a PomZ variant with an intermediate ATPase activity, we took advantage of the synergistic stimulation of PomZ ATPase activity *in vitro* by PomX and PomY. To this end, we analyzed the dynamics of PomX clusters in the absence of PomY. Off-centre as well as midcell clusters showed motion (Fig. 4C); more importantly, MCSD (0.55±0.12μm$^2$ and 0.46±0.10μm$^2$ after 20 min) for both cluster populations and MSD (0.22±0.29μm$^2$ and 0.09±0.09μm$^2$ after 20 min) for off-centre clusters were significantly lower than in WT but still higher than in the PomZ$^{D90A}$ mutant. With the caveats that PomX clusters in the absence of PomY are differently shaped than in the presence of PomY and may not all contain PomZ (Fig. 2C), these data are in agreement with the prediction that reduced ATPase activity reduces cluster motion towards midcell and the mobility of the cluster in general.

We also predicted that perturbing the nucleoid structure would interfere with proper translocation. In *M. xanthus* cells treated for 16 hrs with 50mM hydroxyurea, which inhibits DNA replication, the nucleoid had condensed; however, PomY still formed off-centre and midcell clusters as in untreated cells (Fig. S7FG). Importantly, both cluster populations were less motile than in untreated cells and with significantly reduced MCSD (0.37±0.09µm$^2$ and 0.32±0.08µm$^2$ after 20 min) and MSD (0.26±0.30µm$^2$ and 0.12±0.12µm$^2$ after 20 min) (Fig. 4C).

**Discussion**

Here, we uncover a novel mechanism used by bacteria to identify the incipient division site at midcell. We show that PomX and PomY function together with the ParA/MinD ATPase PomZ to stimulate and position cell division in *M. xanthus* by stimulating Z-ring formation and positioning at midcell. Several lines of evidence support the notion that the complex formed by the PomXYZ proteins directly recruits FtsZ to the division site. First, fewer Z-rings are formed in the absence of any of the three Pom proteins and the few Z-rings formed are randomly localized along the cell length. Second, the PomXYZ complex localizes to the future division site at midcell before and in the absence of FtsZ [here; (Treuner-Lange et al., 2013)]. Third, all three Pom proteins in the PomXYZ complex colocalize with FtsZ at the division site [here; (Treuner-Lange et al., 2013)]. Fourth, PomY and PomZ interact directly with FtsZ. Based on these observations, we conclude that the PomXYZ proteins constitute a novel system for regulation of bacterial cell division and function to mark the incipient division site, recruit FtsZ to this site and stimulate Z-ring formation by FtsZ.

PomZ shares characteristics with other ParA ATPases of the ParA/MinD superfamily. First, PomZ has a low intrinsic ATPase activity. Second, a mutational analysis supports that PomZ dimerizes and binds DNA nonspecifically upon ATP binding, is monomeric in the ADP-bound form and in the apo-form, and spontaneously undergoes ADP-to-ATP nucleotide exchange. Third, ATPase activity is stimulated by nonspecific DNA binding and by AAPs. PomX and PomY independently stimulate ATP hydrolysis by DNA-bound PomZ equally well; however, they function synergistically to stimulate ATP hydrolysis by DNA-bound PomZ suggesting that PomX and PomY may interact differently with PomZ to stimulate ATP hydrolysis. By contrast, ParA, MinD and likely also PpfA in *Rhodobacter sphaeroides* involved in positioning of cytoplasmic chemosensory proteins at midcell (Roberts et al., 2012) have a single AAP, i.e. ParB, MinE and, as suggested for PpfA, TlpT, respectively. To our knowledge, PomZ is the first ParA/MinD ATPase with two distinct AAPs. Interestingly, PomX, PomY, ParB, MinE and TlpT are all non-homologous.

The three Pom proteins interact directly in all pairwise combinations. Strikingly, while expression of different combinations of the Pom proteins in *E. coli* gave rise to different patterns, the simultaneous expression of all three Pom proteins resulted in the formation of nucleoid-associated clusters similar to those observed in *M. xanthus*. Thus, all three Pom proteins are not only required but also sufficient for the formation of the ~15MDa nucleoid-associated complex. The Pom proteins have distinct functions in this complex. In a co-factor independent manner, PomX assembles into filaments *in vitro* and PomY bundles these filaments. In the absence of PomY in *M. xanthus*, PomX assembles into ovoid clusters suggesting that PomX alone also self-assembles *in vivo*. In the presence of PomY, PomX and PomY colocalize and the PomX clusters become more round. *In vivo* this PomXY complex is not associated with the nucleoid, and is stalled randomly in cells. ATP and nucleoid-bound PomZ dimers interact with the PomXY complex and tethers the PomXY complex to the nucleoid and also giving rise to the complex in which PomXYZ colocalize. Notably, the PomXY complex associated with nucleoid-bound PomZ is highly dynamic and translocates across the nucleoid to the midnucleoid at midcell in a biased random walk and undergoes constrained motion at the midnucleoid at midcell. It has previously been pointed out that systems that incorporate a ParA/MinD ATPase use a surface as a scaffold to self-organize (Gerdes et al., 2010; Lutkenhaus, 2012; Vecchiarelli et al., 2012), i.e. MinD associates with the membrane and ParA proteins associate with the nucleoid. Because motion of the PomXYZ cluster and stimulation of PomZ ATPase activity by PomX and/or PomY depends on PomZ DNA binding, we conclude that PomZ uses the nucleoid as a matrix for motion of the PomXYZ cluster.

Based on the recently proposed model for equi-positioning of plasmids and protein complexes (Ietswaart et al., 2014) and our experimental observations, we developed a mathematical model that recapitulates the *in vivo* behavior of the PomXYZ system and explains how the PomXYZ cluster localizes to midcell. In this one-dimensional model, the diffusive flux of nucleoid-bound PomZ into the PomXYZ cluster from either side of the cluster scales with the length of the nucleoid to the left and right side of the cluster. A difference in the two PomZ fluxes into the cluster region results in a local PomZ concentration gradient across the cluster (Fig. 7D). The magnitude of this concentration gradient scales with the asymmetry of the cluster position on the nucleoid (Fig. 7D). Assuming that the cluster preferentially moves up the PomZ gradient then the PomXYZ cluster will have a bias for translocation in the direction facing most of the nucleoid. If the cluster is at the midnucleoid, which coincides with midcell until the chromosomes have segregated, the PomZ flux is equal from either side of the cluster, therefore there will be no PomZ concentration gradient over the cluster, no bias in the movements, and the motion of the cluster, as observed experimentally, is constrained to midcell. Of note, should the cluster overshoot in one direction and leave the midcell area, then this system is self-correcting and will eventually bring the cluster back to the midnucleoid at midcell. Thus, in this model, the difference in the diffusive PomZ flux into the cluster from either side is a proxy for PomXYZ cluster asymmetry on the nucleoid and converts the global intracellular asymmetry of the PomXYZ cluster into a local PomZ concentration gradient over the cluster and, therefore, biased or unbiased motion.

The systems that regulate cell division in growing bacteria are diverse. The PomXYZ system integrates features from these divergent systems in a novel fashion. Similar to MapZ, PomXYZ localize to the future division site before as well as in the absence of FtsZ, colocalize with FtsZ at the division site and are important for Z-ring formation and positioning. On the other hand, MapZ is translocated by cell growth to the future division site whereas the PomXYZ complex self-organizes to actively translocate to the midnucleoid at midcell. Similarly, to the Min systems in *E. coli* and *B. subtilis* and the MipZ/ParB system, the PomXYZ system incorporates a ParA/MinD ATPase. On the other hand, the *E. coli* Min system self-organizes to undergo coupled pole-to-pole oscillations on the membrane, the *B. subtilis* Min system localizes to the cell poles and septum, and the MipZ/ParB system self-organizes to form bipolar MipZ gradients over the nucleoid and all three systems inhibit Z-ring formation throughout cells except at midcell. By contrast, the PomXYZ system self-organizes to establish a cellular pattern in which the complex translocates over the nucleoid by a biased random walk to the midnucleoid at midcell and undergoes constrained motion at midnucleoid at midcell where it directly stimulates Z-ring formation. It will be interesting to address the physiological, ecological and evolutionary advantages of these different systems involved in the spatiotemporal regulation of cell division in bacteria. Nonetheless, we predict that the PomXYZ system would be functional independently of the mode of cell growth as well as cell size and shape and would only depend on a centrally localized and segregating nucleoid.

ParA/MinD ATPases are highly versatile and together with their AAP(s) can self-organize giving rise to different patterns within cells using the membrane or the nucleoid as a scaffold. We suggest that tuning of the interactions between a ParA/MinD ATPase, its cognate AAP(s) and scaffold ultimately results in different patterns formed including pole-to-pole oscillations, bipolar gradients and, as shown here, a biased random walk to midnucleoid and constrained motion at midnucleoid. In the case of the cytoplasmic cluster of chemosensory proteins in *R. sphaeroides*, this cluster also localizes over the midnucleoid at midcell and then splits into two clusters that are distributed to the daughters (Roberts et al., 2012). In this system, the ParA ATPase (PpfA) localizes symmetrically on either side of the protein cluster as well as with the cluster of chemosensory proteins. In the cyanobacterium S*ynechococcus elongatus* PCC7942, a ParA protein is important for correct positioning of carboxysomes along the cell length (Savage et al., 2010). While the details of this system are not known, we speculate that this system as well as the system that positions chemosensory proteins at midcell may function by a mechanism as described here for the PomXYZ system and for plasmid and protein complex positioning over the nucleoid (Ietswaart et al., 2014).

Systems that regulate Z-ring positioning and in that way cell division generally couple chromosome replication and segregation to cell division to ensure that each daughter cell receives the correct chromosome complement. In the case of the PomXYZ system, our data suggest that PomY as well as PomZ in the PomXYZ complex directly interact with FtsZ to stimulate Z-ring formation and positioning. However, the PomXYZ complex colocalizes over the midnucleoid at midcell with the Z-ring for several hours before replication and segregation are complete and cell constriction initiates. As in other bacteria, the precise cue that triggers the cytokinetic machinery including FtsZ to initiate constriction in *M. xanthus* is not known. Similarly, we do not know why the PomXYZ complex only recruits FtsZ at the midnucleoid at midcell and not during the translocation from an off-centre position to the midnucleoid. These questions will be addressed in future experiments.

**Experimental Procedures**

Details on strain construction and growth conditions are described in Supplemental Experimental Procedures. *M. xanthus* strains, plasmids and primers are listed in Tables S3, S4 and S5, respectively. FRAP experiments, immunoblot analysis, yeast two hybrid assays, protein purification, EM, protein sedimentation assay and ATPase assay were done as detailed in Supplemental Experimental Procedures.

Light microscopy. Cells from exponentially growing cultures were transferred to slides containing a thin pad of 1% SeaKem LE agarose (Cambrex) with TPM buffer (10mM Tris-HCl pH 7.6, 1mM $KH_2PO_4$ pH 7.6, 8mM $MgSO_4$), covered with a coverslip and imaged with a temperature-controlled Leica DMI6000B microscope with adaptive focus control, a motorized stage (Prior) and a HCX PL APO 100x/1.47 oil Corr TIRF objective at 32°C with a Hamamatsu Orca Flash 4.0 using Leica MM AF software. Image processing and data analysis was performed using Metamorph® v 7.5 (Molecular Devices). For DAPI staining, cells were incubated with 1µg/ml DAPI for 10 min at 32°C prior to microscopy. For time-lapse recordings, cells were placed on a TPM-buffered agarose pad containing 0.2% CTT medium containing TetraSpeck™ 0.5µm fluorescently labelled beads (Molecular Probes™) for image alignment and analysis of cluster dynamics. Automated picture alignment and semi-automated cluster tracking was done using Metamorph® v 7.5. For tracking cluster dynamics, the maximum identity method was used with a minimal identity of 60% and a maximal displacement of 25 pixels. Kymographs were generated using Metamorph® v 7.5.

**Acknowledgement**

We thank Susan Schlimpert, Martin Thanbichler, Anke Treuner-Lange, Emanuel Reithmann and Jacob Halatek for many helpful discussions. This work was supported by the German Research Council (DFG) within the framework of the Transregio 174 "Spatiotemporal dynamics of bacterial cells", the Max Planck Society and the Graduate School of Quantitative Biosciences, Munich.

**Author contributions**
Conceived and designed experiments: DS, LS-A
Developed theoretical concepts and models: EF, SB
Performed experiments: AH, DS, JV
Analyzed data: DS, EF, LS-A, SB
Contributed reagents/materials/analysis tools: SH
Wrote paper: DS, EF, LS-A, SB

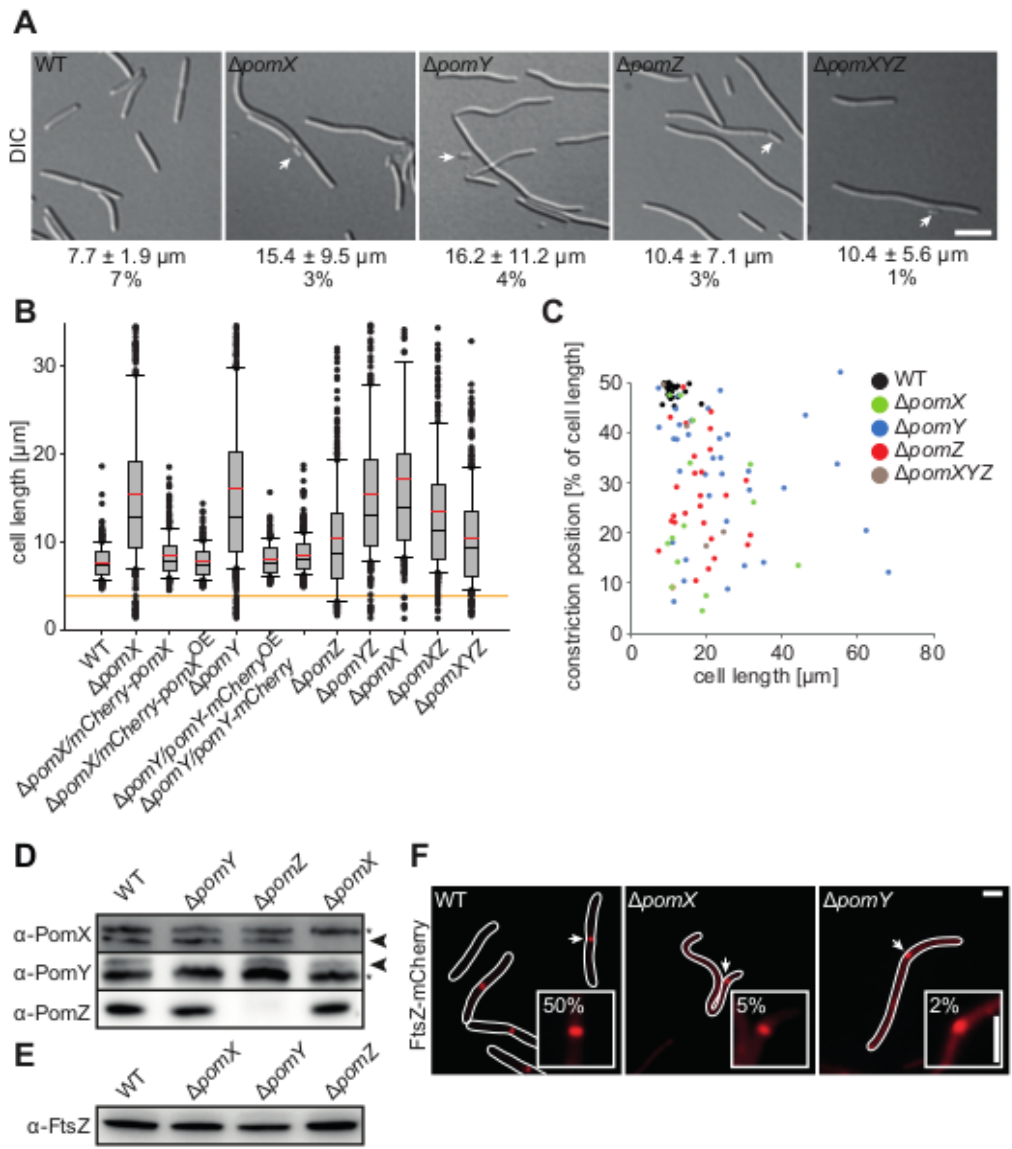

Figure 1

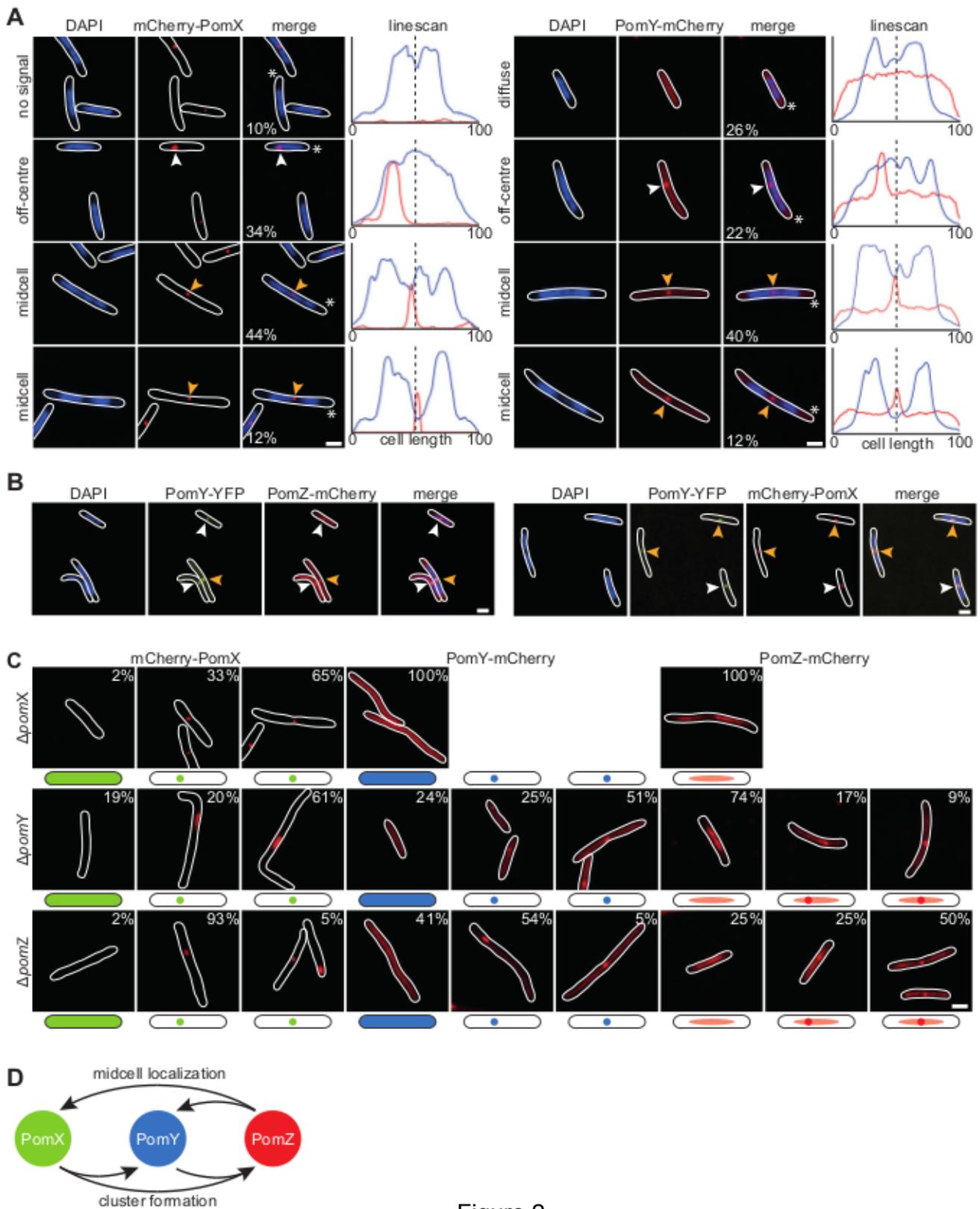

Figure 2

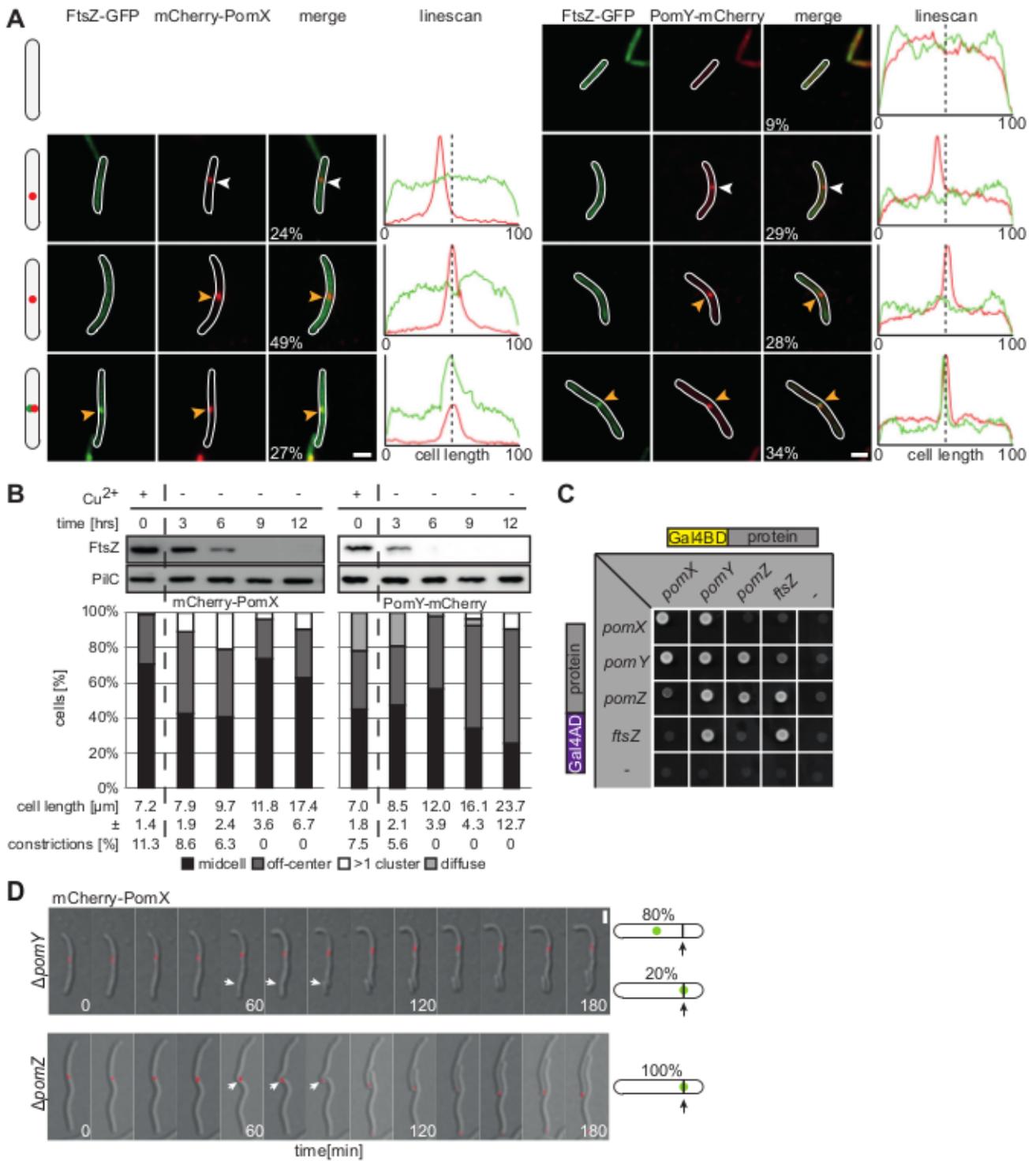

Figure 3

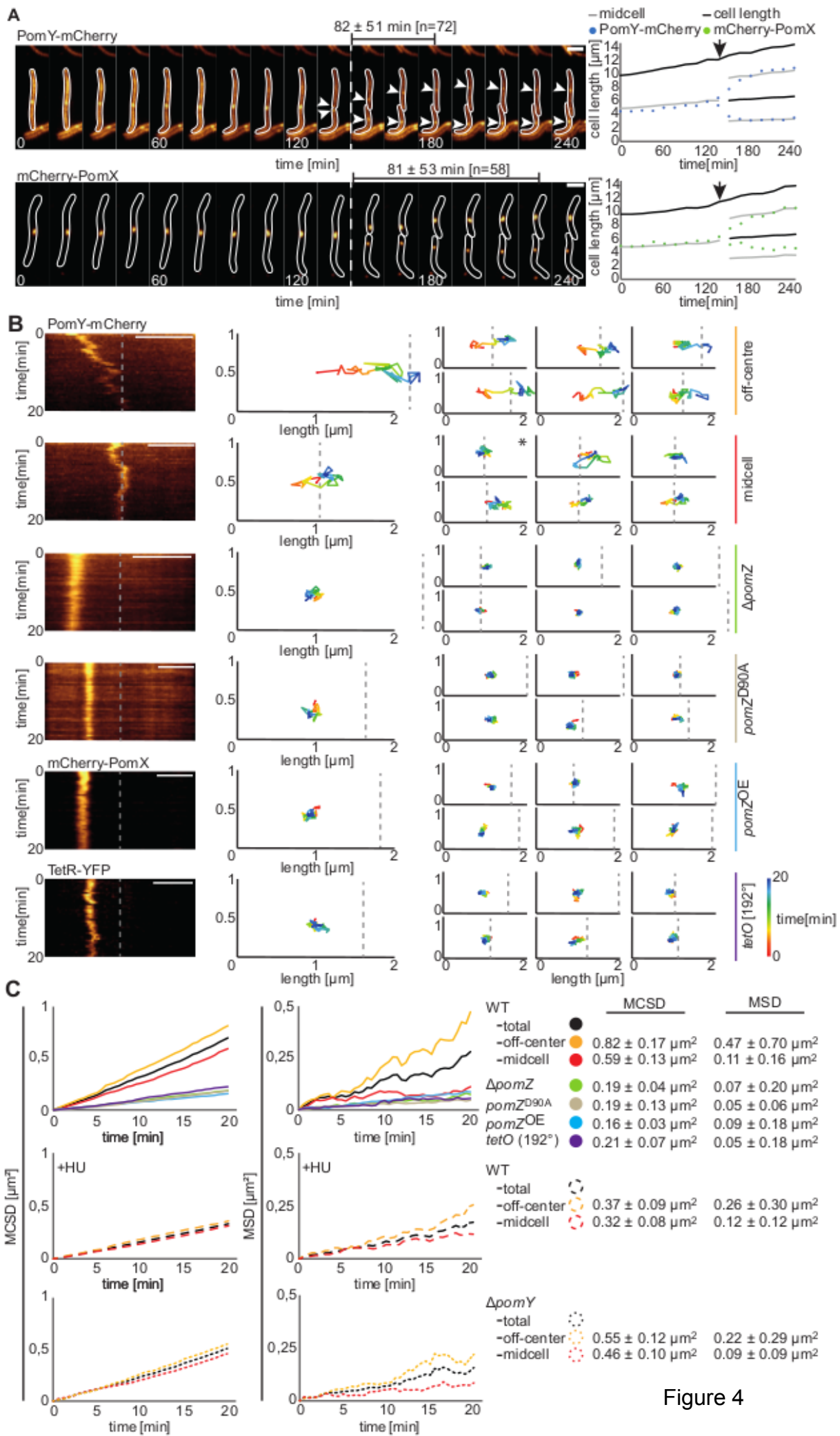

Figure 4

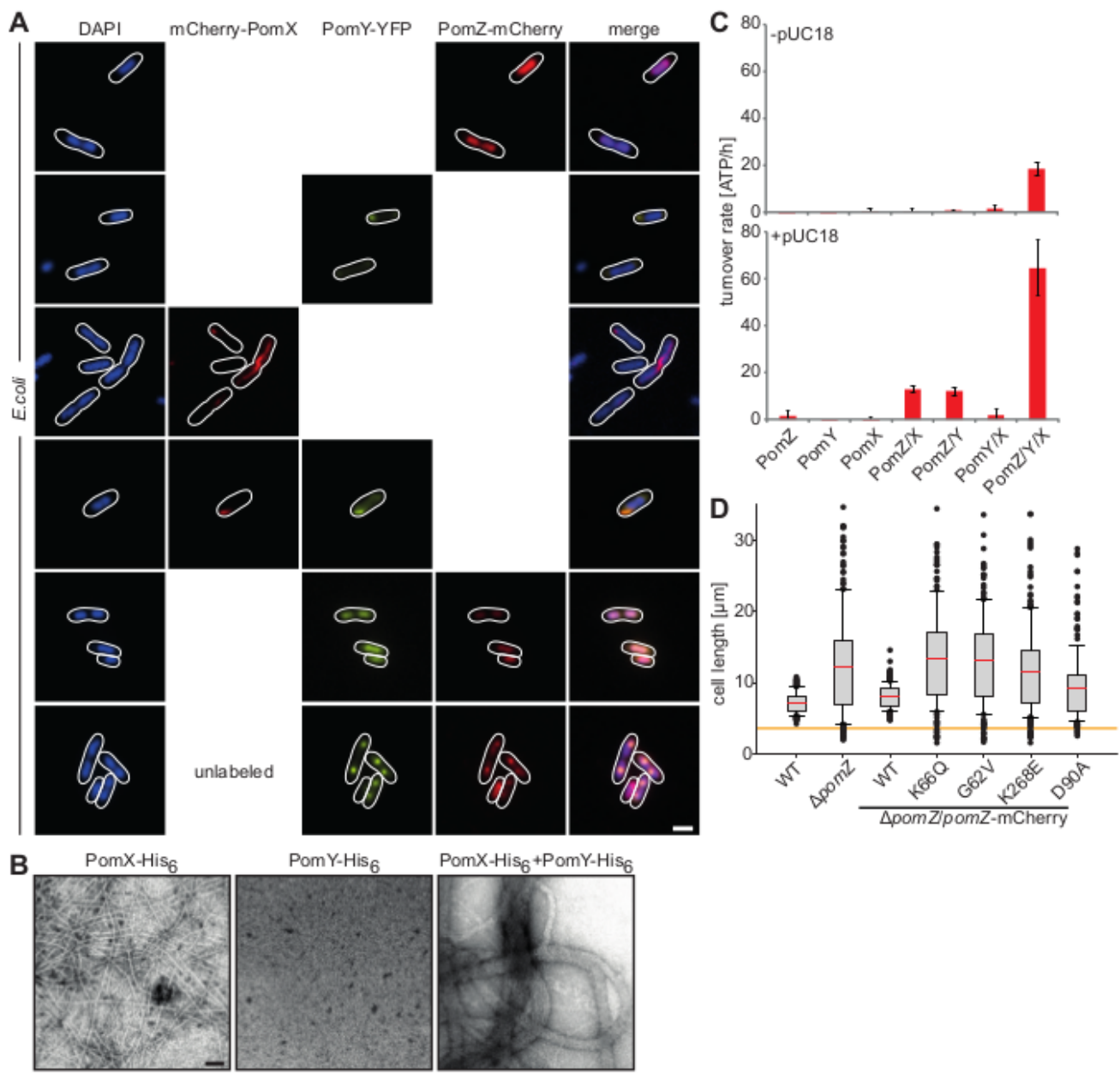

Figure 5

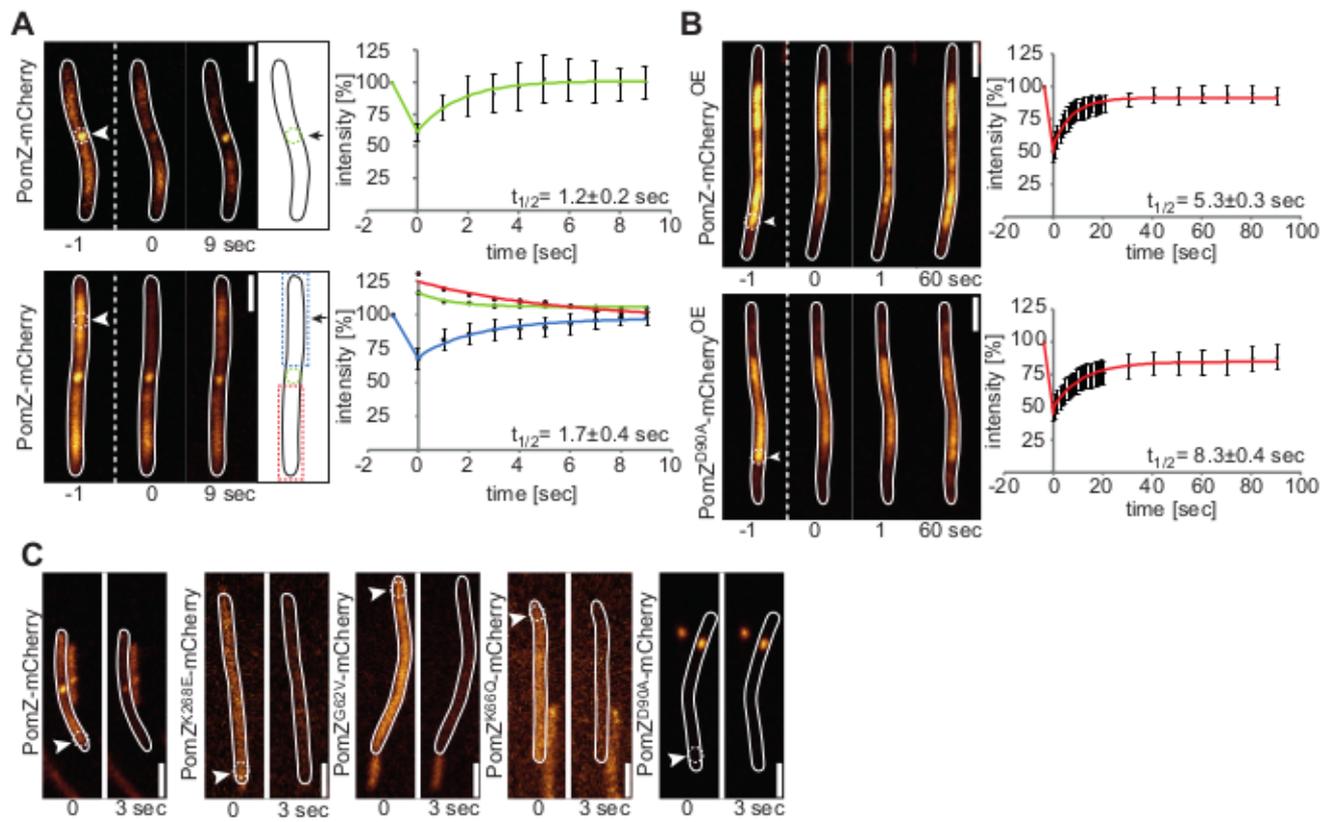

Figure 6

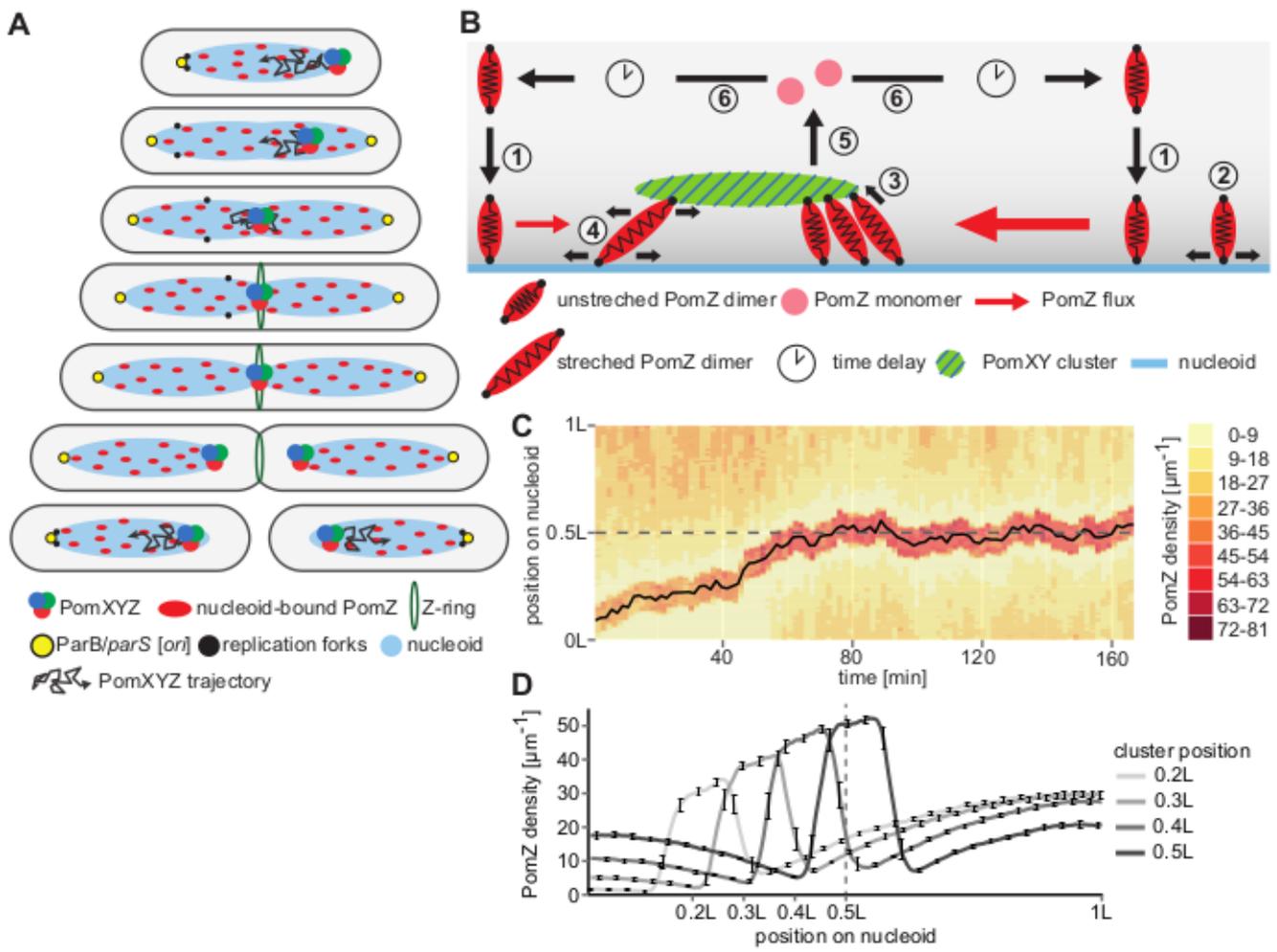

Figure 7

**Legends to Figures**

Figure 1: PomX and PomY are important for cell division and Z-ring formation and positioning
A. Morphological changes in cells of the indicated genotypes. Arrows indicate minicells. Numbers indicate mean cell length ± standard deviation (SD) and constriction frequency. Scale bar, 5µm.
B. Box plots of cell length distributions of strains of the indicated genotypes (n>200). Boxes enclose the 25$^{th}$ and 75$^{th}$ percentile with red lines representing the mean and whiskers the 10$^{th}$ and 90$^{th}$ percentile. Note that the few cells longer than 35µm are not included in the plots. Dots below the orange line indicate minicells, smaller than 4.6µm. The strains labelled mCherry PomX$^{OE}$ and PomY-mCherry$^{OE}$ overexpress the two proteins.
C. Lack of PomX, PomY or PomZ leads to misplaced cell divisions. Each dot represents a cell division constriction plotted as percent of cell length and as a function of cell length in strains of indicated genotypes.
D. Immunoblot analysis of the accumulation of PomX, PomY and PomZ. Equal amounts of total protein from cells of the indicated genotypes were applied. * indicate cross-reacting proteins and arrows PomX and PomY.
E. Immunoblot analysis of FtsZ accumulation. Equal amounts of protein from cells of the indicated genotypes were loaded per lane.
F. PomX and PomY are important for Z-ring formation and positioning. $ftsZ^+$ cells with the indicated mutations, expressing FtsZ-mCherry, were visualized. Strains used from left to right: SA3139, SA4228, SA4707. Arrows indicate Z-rings shown at a higher magnification in the insets. Numbers represent Z-ring frequencies (n>100). Scale bars, 2µm.
See also Figures S1 and S2.

Figure 2. PomX, PomY and PomZ localize similarly and form a complex that is positioned at midcell by PomZ
(A, B). PomX, PomY and PomZ have similar localization patterns and colocalize. mCherry-PomX, PomY-mCherry and PomZ-mCherry were localized in the corresponding in-frame deletion mutants (n>200). White and orange arrows indicate off-centre and midcell clusters. DAPI was used to stain nucleoids. In A, numbers indicate % of cells with that localization pattern. Linescans show the fluorescence intensity of DAPI (blue) and mCherry-PomX/PomY-mCherry (red) fluorescence as a function of normalized cell length for cells marked with *. Midcell is indicated by the stippled line. Midcell clusters are divided into those colocalizing with the nucleoid at midnucleoid and those localized between two segregated nucleoids (lower panels). Strains used in A from left to right: SA4229, SA4713 (n>200). In B, PomY-YFP expression was induced by 150µM Cu$^{2+}$. Strains used from left to right: SA7020, SA7041.
C. PomX, PomY and PomZ localize interdependently. The fusion proteins were analyzed in the in-frame deletion mutants listed on the left. Numbers indicate % of cells with that localization (n>200). Cartoons illustrate localization patterns schematically with mCherry-PomX in green, PomY-mCherry in blue and PomZ-mCherry in red. Strains used in A from left to right: Top row SA4252, SA4737, SA4232; middle row SA4739, SA4712, SA4706; bottom row SA5821, SA4720, SA3131.
D. Schematic of localization dependency of PomX, PomY and PomZ.
Scale bar, 2µm in all panels.

See also Figures S1 and S3.

Figure 3. PomX and PomY localize at midcell before and in the absence of FtsZ
A. mCherry-PomX and PomY-mCherry localize at midcell before FtsZ. mCherry-PomX and PomY-mCherry were expressed in the presence of 150µM $Cu^{2+}$. Three localization patterns in the case of mCherry-PomX (left panels) and four in the case of PomY-mCherry (right panels) were observed with the indicated percentages (n>200). Linescans are as in Fig. 2A with FtsZ (green) and mCherry-PomX/PomY-mCherry (red). Cartoons on the left indicate localization patterns of FtsZ (green) and PomX/PomY (red). White and orange arrows indicate off-centre and midcell clusters, respectively. Scale bar, 2µm. Strains used from left to right: SA4295, SA4736.
B. PomX and PomY localize at midcell in the absence of FtsZ. FtsZ was expressed from a $Cu^{2+}$ inducible promoter in the presence of 300µM $Cu^{2+}$. FtsZ was depleted by removal of $Cu^{2+}$ from the growth medium (t = 0 hrs) and samples withdrawn at the indicated time points. Upper panel, FtsZ levels during $Cu^{2+}$ depletion. Lower panel, the PilC loading control in the same cells. For each time point, cells (n>150) were analyzed for cell length ± SD, constriction frequency and localization of mCherry-PomX/PomY-mCherry. Pattern frequency is shown in bar graphs in % of total cells. Strains used from left to right: SA5809, SA4718.
C. Yeast two hybrid analysis for interactions between PomX, PomY, PomZ and FtsZ. Yeast strain AH109 was co-transformed with pairs of plasmids encoding the indicated variants of Gal4-AD and Gal4-BD. Interactions are evidenced by AH109 growth on selective medium. Negative control (-): Co-transformation of AH109 with a bait plasmid containing a Gal4-AD or Gal4-BD fusion and a plasmid expressing the native Gal4-AD or Gal4-BD protein.
D. PomY is required to align cell divisions with PomX clusters. Cells of the indicated genotypes expressing mCherry-PomX were followed by time-lapse microscopy for 180 min on an agarose pad containing 0.25% CTT growth medium at 32°C. Images were recorded every 15 min. Images are merged DIC and fluorescence microscopy images. White arrows indicate division constrictions. Right, cartoons represent localization of constrictions (black arrow) relative to the mCherry-PomX clusters (green). Numbers display frequency of indicated localization patterns (n>25 cells per analysis). Strains used top to bottom: SA7008, SA7009.
See also Figures S1 and S4.

Figure 4. PomX and PomY form dynamically localized clusters that are positioned at midcell by PomZ
A. mCherry-PomX and PomY-mCherry are dynamically localized. Time-lapse microscopy of cells expressing mCherry-PomX or PomY-mCherry done as in Fig. 3D. Numbers above the images indicate the mean ± SD translocation time from the release of a cluster at a cell division site until it reached the new midcell. Black lines indicate this translocation time for the cells shown. Right, schematics illustrate cluster localization in the cells on the left. The stippled line in the time-lapse indicates a cell division that is marked by a black arrow in the schematic. White arrows mark PomY-mCherry clusters. Scale bar, 2µm. Strains used (top to bottom): SA7000, SA4797.
B. Kymographs of PomY-mCherry, mCherry-PomX or TetR-YFP translocation. Time interval between image capture was 30 sec for 20 min. Right panels, representative two-dimensional cluster trajectories color-coded for time. Stippled lines indicate midcell. Large panels refer to the clusters show in the kymographs. Scale bars, 2µm. TetR-YFP was expressed in presence of 150µM vanillate. * mark a cell in which the midcell cluster was essentially non-motile. Strains used from top to bottom: SA4746 (two top panels), SA4796, SA7027, SA7022, SA6757.

C. Quantification of PomY-mCherry, mCherry-PomX and TetR-YFP cluster translocation. Cluster centroids were tracked for >30 cells per strain and used to calculate MCSD (left panel) and MSD (right panel). For analysis of dynamics after hydroxyurea treatment, exponentially growing cells were treated with 50mM hydroxyurea for 16 hrs at 32°C before fluorescence microscopy. Colors as in B.

See also Figures S5 and S7.

Figure 5. PomX, PomY and PomZ interact in all pairwise combinations and PomXY stimulate PomZ ATPase activity

A. PomX, PomY and PomZ self-assemble to form clusters that colocalize with the nucleoid in *E. coli*. Three top rows, PomZ-mCherry, mCherry-PomX and PomY-YFP expression was induced in *E. coli* BL21 DE3 in the presence of 0.05mM IPTG for 1 hr at 37°C. For PomZ-mCherry/PomY-YFP or mCherry-PomX/PomY-YFP co-expression, PomZ-mCherry and mCherry-PomX expression was induced as described and PomY-YFP by addition of 0.015% arabinose. For PomZ-mCherry/PomY-YFP/PomX co-expression, PomZ-mCherry and PomY-YFP expression was induced for 1 hr with 0.05mM IPTG before PomX expression was induced with 0.015% arabinose for 30 min at 37°C. Cells were grown in LB medium containing 0.2% glucose and treated with chloramphenicol for 30 min to condense the nucleoids before DAPI staining and fluorescence microscopy. Scale bar, 2µm.

B. PomX-His$_6$ forms filaments that are bundled by PomY-His$_6$ *in vitro*. EM images of negatively stained PomX-His$_6$ (final concentration 3µM) and PomY-His$_6$ (final concentration 3µM) alone and after mixing in a 1:1 molar ratio of 3µM final concentration each. Scale bar, 100nm.

C. His$_6$-PomZ ATPase activity is stimulated by the PomXY complex. Specific ATPase activity of His$_6$-PomZ (final concentration 2µM) was measured in the presence of 1mM ATP alone, with or without PomX-His$_6$, PomY-His$_6$ (final concentration 2µM each) and 5nM pUC18 plasmid DNA. Experiments were performed in triplicates and results shown as mean ± SD.

D. PomZ variants affected in the ATPase cycle do not correct the cell division defects in a Δ*pomZ* mutant. Box plots are as in Fig. 1B. Strains used from left to right: DK1622, SA3108, SA3131, SA5001, SA5000, SA5837, SA3146.

See also Figures S4 and S6.

Figure 6. PomZ is rapidly exchanged in the PomXYZ cluster and diffuses rapidly on the nucleoid

A. FRAP analysis on cells expressing PomZ-mCherry. A 5 pixel circular region (white stippled circles) on the PomZ-mCherry cluster (upper panel) or on the nucleoid outside of the cluster (lower panel) was photo-bleached, and recovery was followed by fluorescence microscopy. Images were recorded every second. Scale bar, 2µm. Stippled lines indicate the photo-bleaching event. Cartoons show the areas used for recovery measurements in stippled colored frames and graphs show the average relative integrated intensities in these areas as a function of time. The recovery half-time ($t_{1/2}$) was determined by fitting the mean data to a single-exponential function (n=20). Strain used: SA3131.

B. FRAP analysis on cells overexpressing PomZ-mCherry variants. Δ*pomZ* cells overexpressing (OE) PomZ-mCherry or PomZ$^{D90A}$-mCherry were subjected to photo-bleaching in a 5 pixel circular region (white stippled circles). Stippled line indicate photo-bleaching event. Recovery in the bleached area was followed for 100 sec. Images were acquired every 300 msec for 20 sec and then every 5 sec. Graphs represent the average relative integrated intensities of the bleached region as a function of time. The recovery half-time ($t_{1/2}$) was calculated as in A (n=20). Scale bar, 2µm. Strains used from top to bottom: SA7011, SA4799.

C. Photo-bleaching experiments with PomZ-mCherry variants. Δ*pomZ* cells expressing PomZ-mCherry variants were exposed to photo-bleaching for 3 sec in a 5 pixel circular region (white stippled circles) on the nucleoid outside of the cluster in the case of PomZ$^{WT}$ and PomZ$^{D90A}$ and on the nucleoid in the case of the three remaining strains. Scale bar, 2µm. Strains used from left to right: SA3131, SA5837, SA5000, SA5001, SA3146.

Figure 7. A PomZ flux-based mechanism for midcell positioning of the PomXYZ complex

A. The PomXYZ complex is dynamically localized on the nucleoid. Schematic illustrates localization of the complex during the cell cycle starting with a cell immediately after cell division (top). The trajectories indicate the imminent biased random motion of off-centre complexes towards midcell and constrained motion at midcell.

B. Schematic of the PomZ flux-based model. See main text for details.

C. Kymograph showing a representative simulation of PomZ localization on the nucleoid. The parameters listed in Table S2 were used in the simulation. The PomZ distribution on the nucleoid is averaged over time in intervals of about 100 sec and plotted against time. Right, color code for PomZ density. Black line, trajectory of the midpoint of the cluster over time. L, nucleoid length and dashed grey line midnucleoid.

D. Average density of PomZ on the nucleoid for different cluster positions. The PomZ density profile on the nucleoid (as shown in C over time for one run) were averaged when the cluster passed 0.2L, 0.3L, 0.4L and 0.5L (midnucleoid, dashed grey line) using 100 runs of the stochastic simulation. Error bars indicate 95% confidence intervals and are shown for 20% of the average density values.

See also Figure S7.

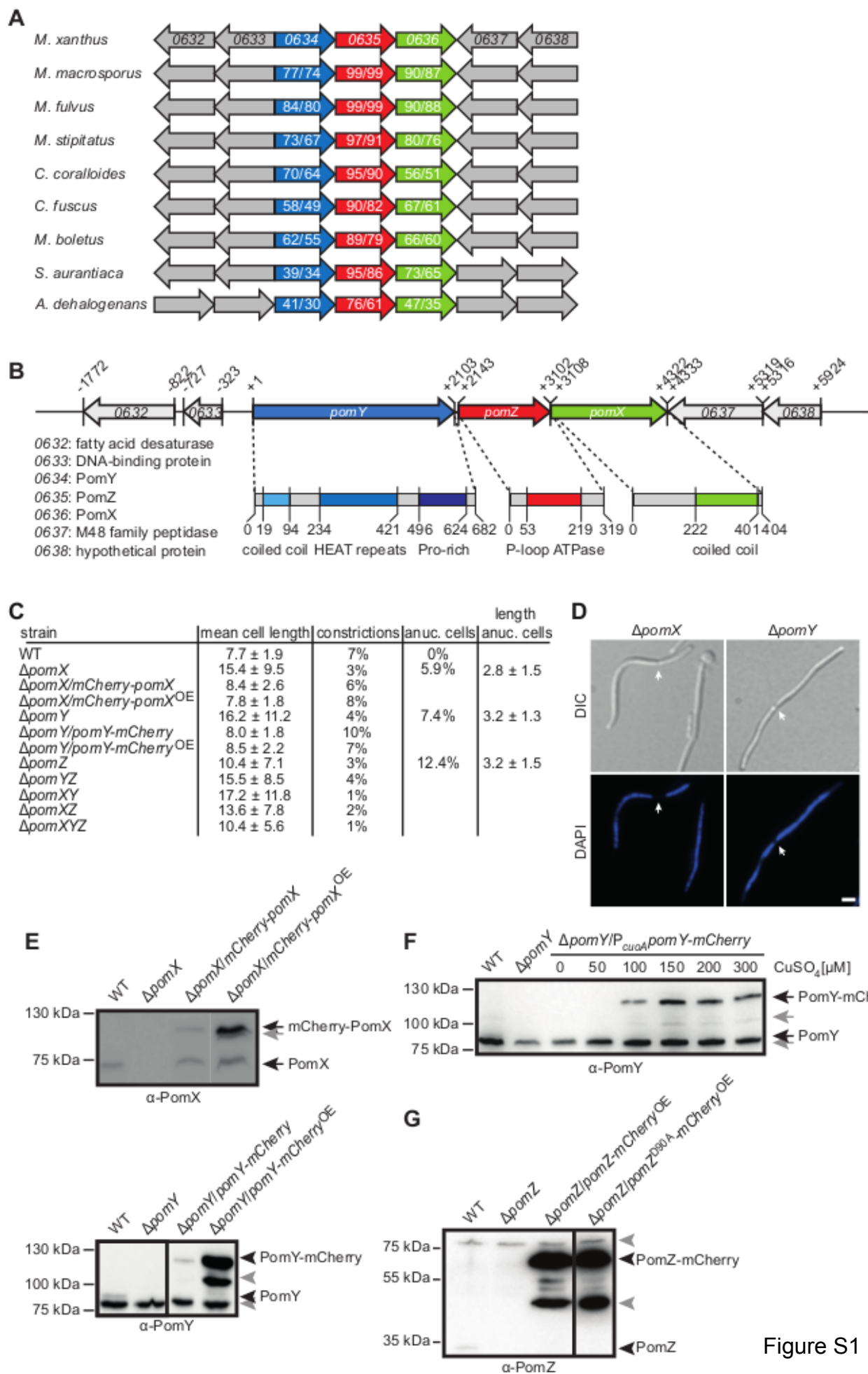

Figure S1

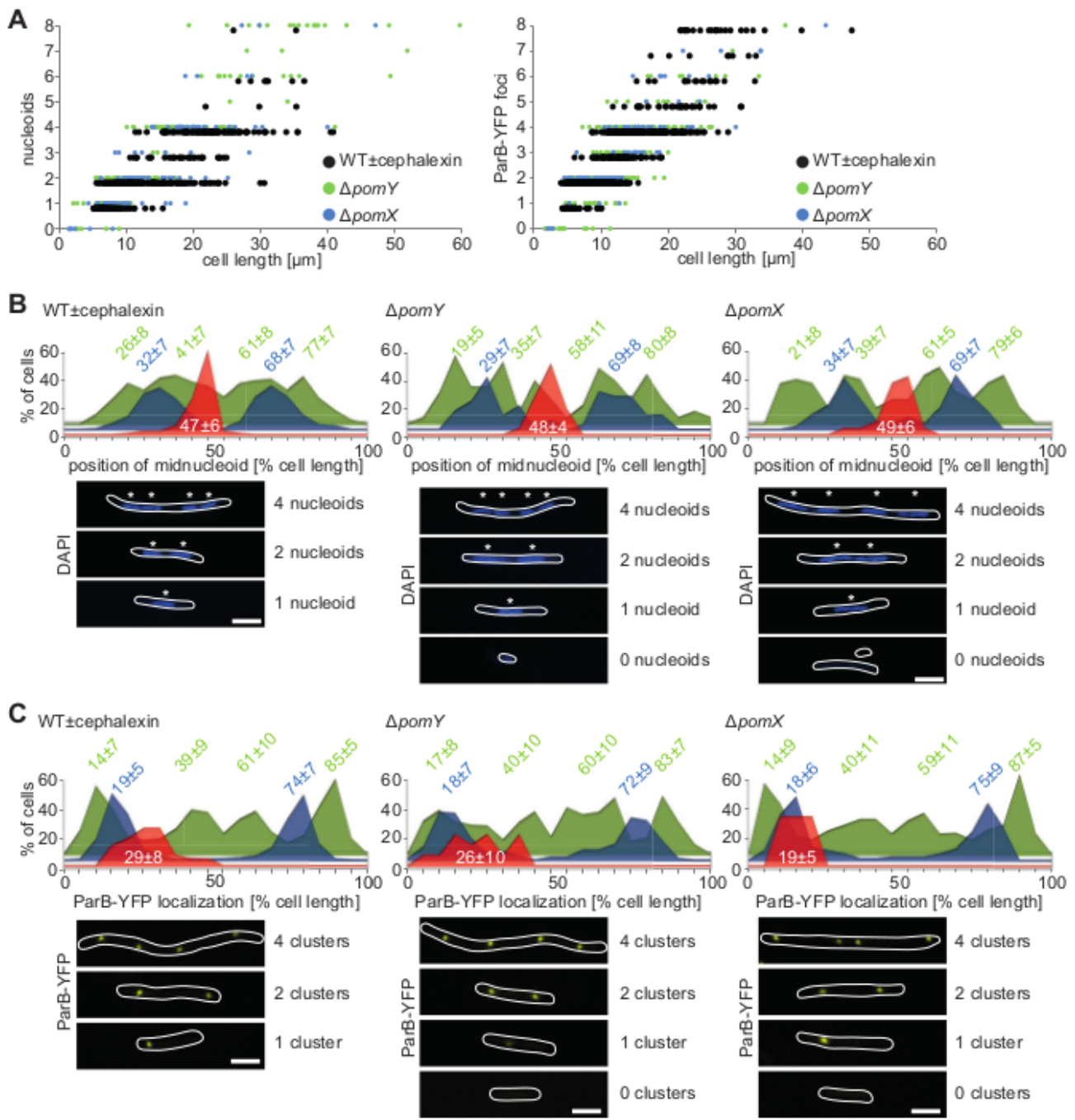

Figure S2

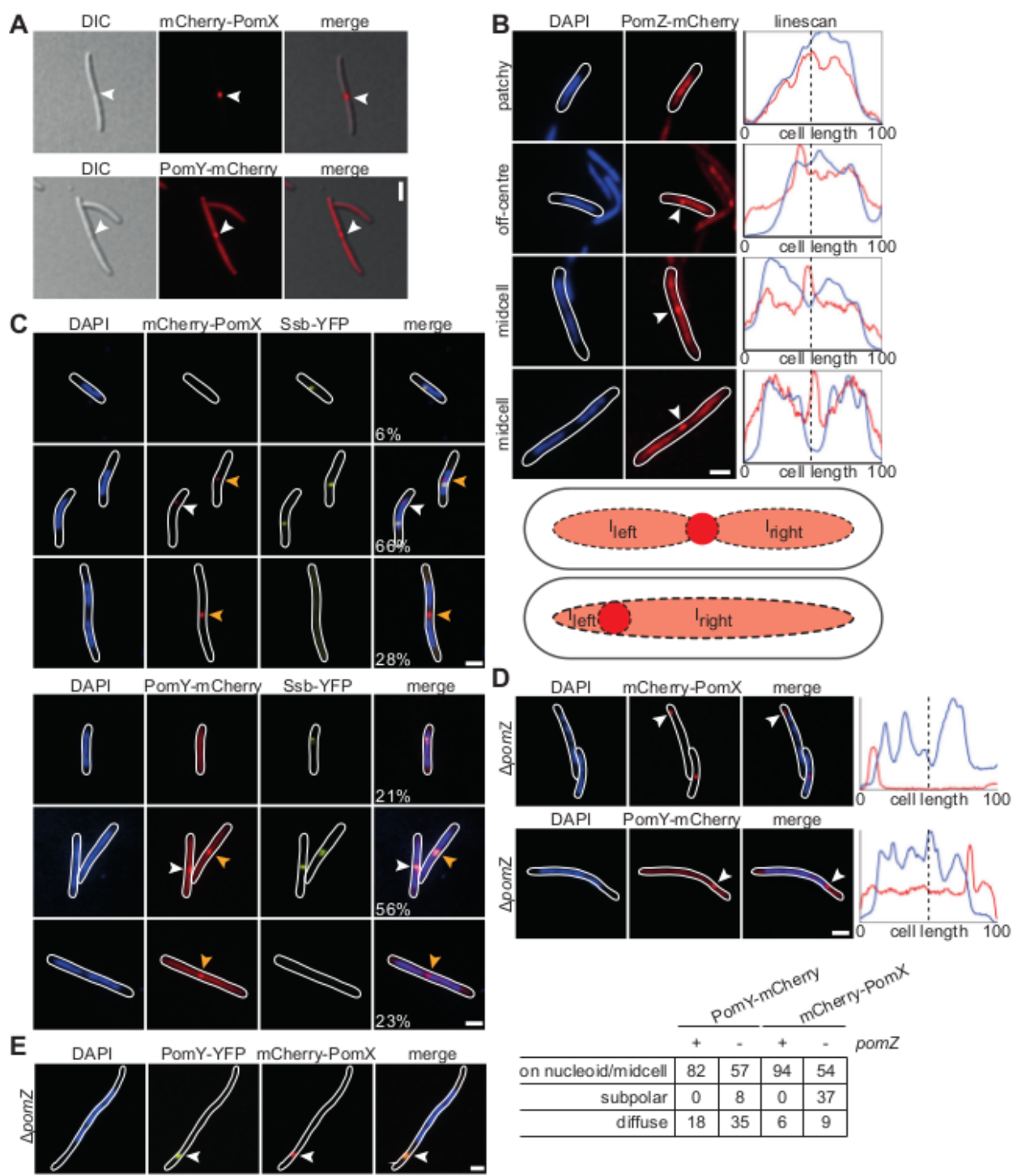

Figure S3

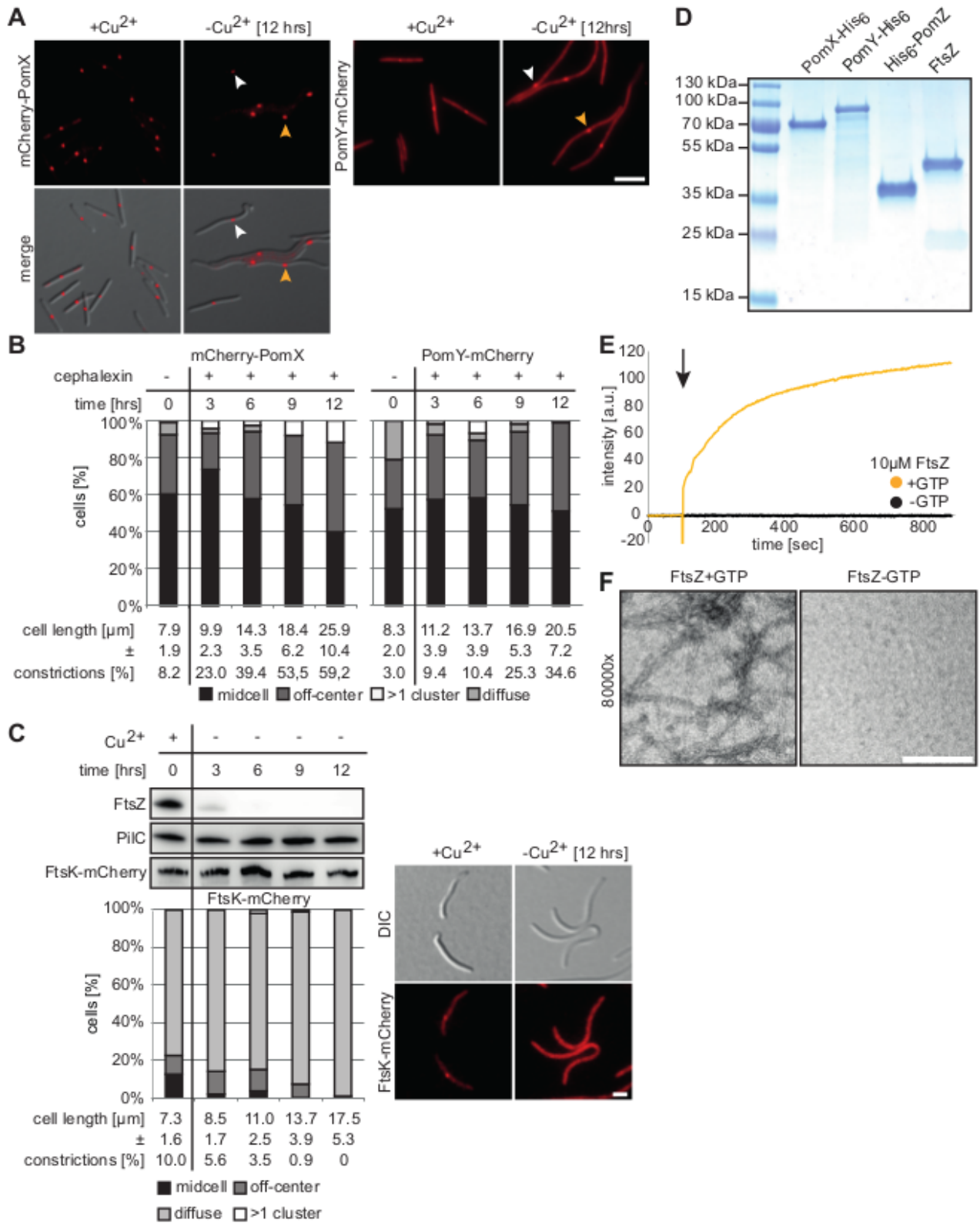

Figure S4

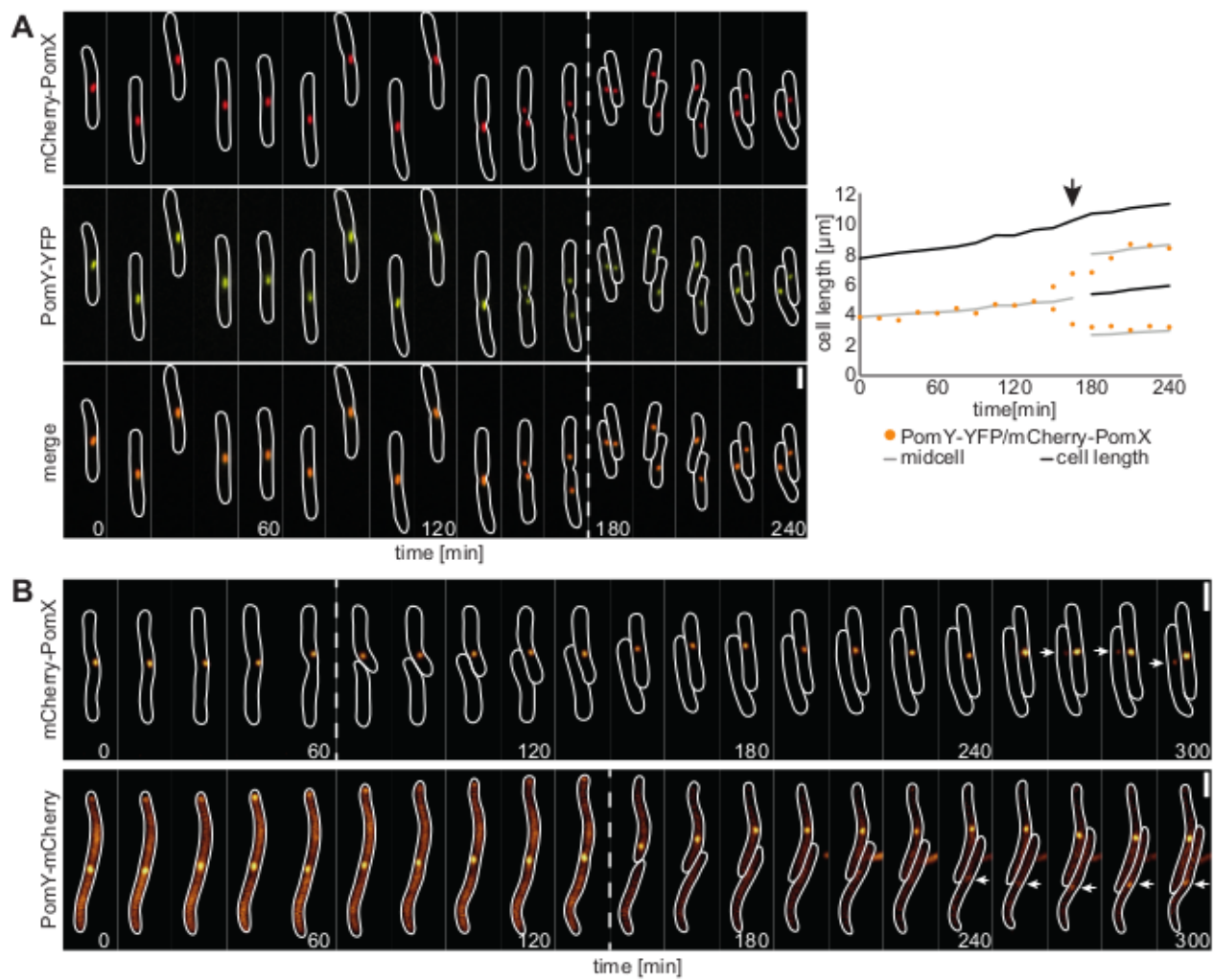

Figure S5

Figure S6

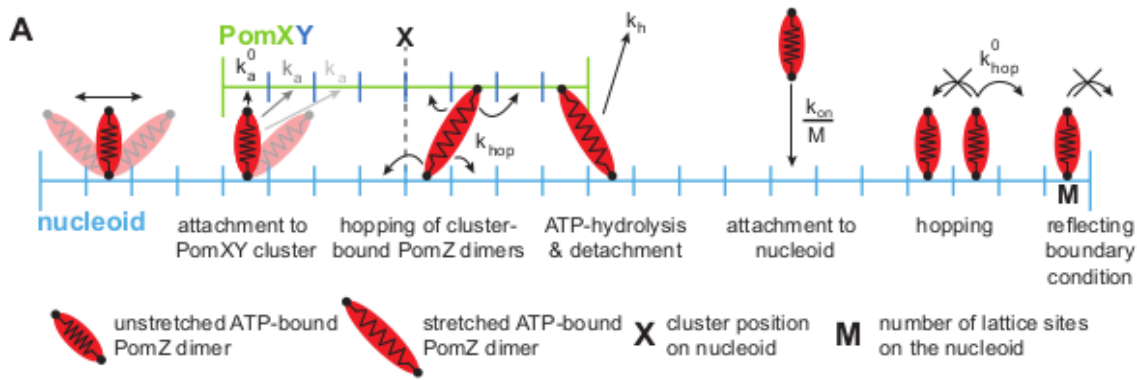

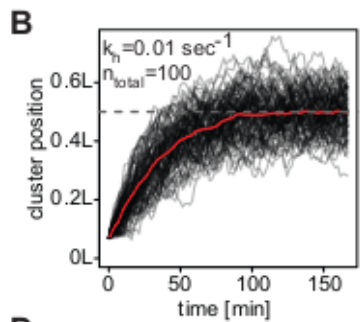 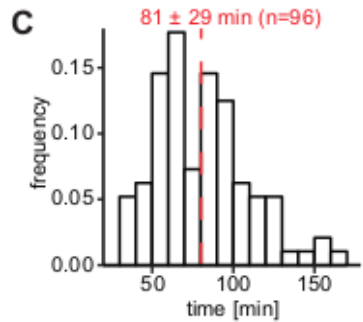

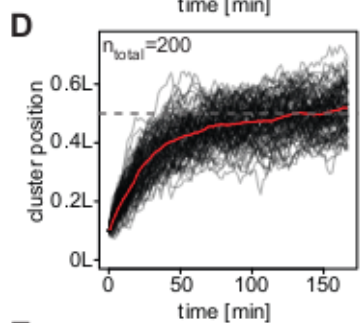 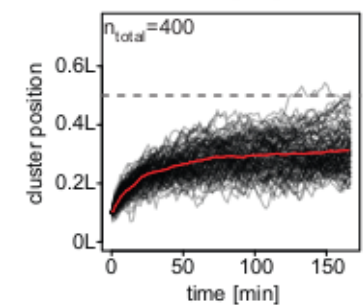 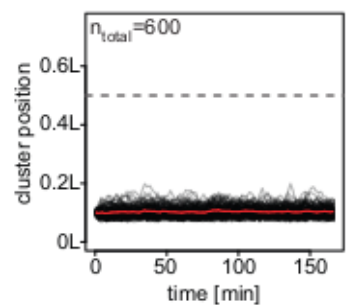

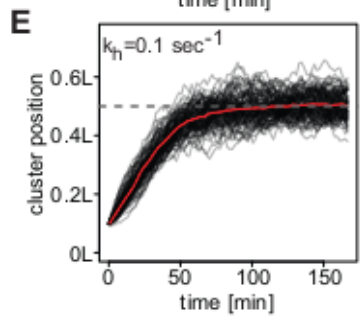 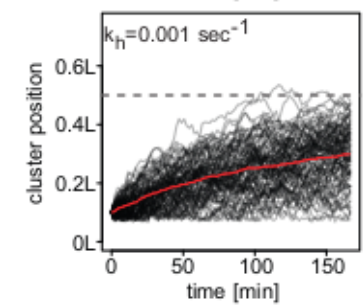 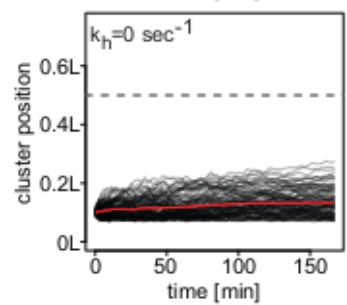

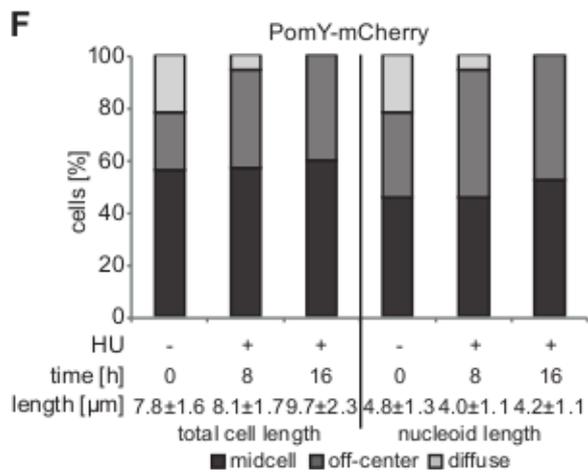 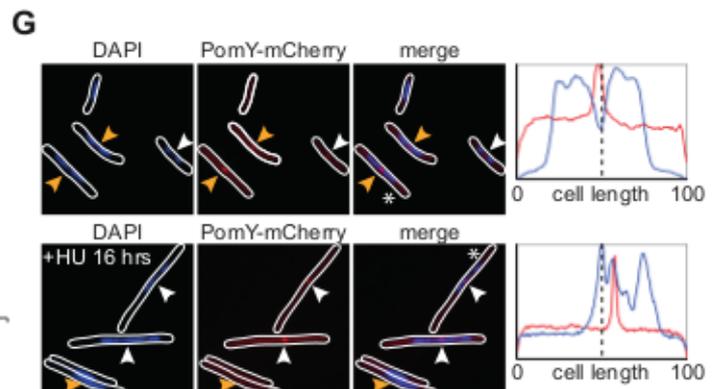

Figure S7

# Supplemental Information

**The PomXYZ proteins self-organize on the bacterial nucleoid to stimulate cell division**

Dominik Schumacher, Silke Bergeler, Andrea Harms, Janet Vonck, Sabrina Huneke,

Erwin Frey & Lotte Søgaard-Andersen

**This file contains:**

- Supplemental Information: Theory
- Supplementary Experimental Procedures
- Legends to Supplementary Figures S1-S7
- Supplementary Tables S1-S5
- Supplemental References

**Supplemental Information: Theory**

**A flux-based model for motion and positioning of the PomXYZ cluster**

We searched computationally for mechanisms that could give rise to a biased random walk of the PomXYZ complex to midnucleoid and constrained motion over the midnucleoid. We propoe a model that is based on the following experimental observations and inferences:

**1.** PomX, PomY and PomZ are required and sufficient for nucleoid-associated cluster formation (Fig. 5A). PomX nucleates the formation of this cluster and PomX and PomY independently of PomZ assemble to form a cluster (Fig. 2C; S3E). PomZ recruits this cluster to the nucleoid and is central to localization of the cluster at midnucleoid at midcell (Fig. 2C; S3D).

**2.** PomZ in its ATP-bound dimeric form binds nonspecifically to the nucleoid (Fig. 5A, S6CD), interacts with the PomXY complex and recruits it to the nucleoid (Fig. S3DE).

**3.** Only nucleoid-bound dimeric PomZ interacts with the PomXY complex (Fig. S6DFG).

**4.** PomZ alone whether bound to DNA or not has a low ATP turnover rate (Fig. 5C). PomX and PomY together stimulate the ATPase activity of PomZ in the presence of DNA (Fig. 5C) suggesting that PomZ can simultaneously interact with PomXY and DNA.

**5.** Detachment of a PomZ dimer from the nucleoid depends on ATP hydrolysis (Fig. 6BC). Therefore, detachment of PomZ from the nucleoid occurs primarily at the PomXYZ cluster and PomZ in the PomXYZ cluster exchanges rapidly in an ATP hydrolysis-dependent manner (Fig. 6AC). After ATP hydrolysis, ADP-bound PomZ monomers detach from the nucleoid and the cluster. Before PomZ molecules can rebind to the nucleoid, they undergo ADP-to-ATP exchange and form dimers (Fig. S6BD).

**6.** PomZ bound to the nucleoid undergoes fast diffusion on the nucleoid giving rise to a diffusive PomZ flux on the nucleoid (Fig. 6ABC).

**7.** PomZ undergoes fast diffusion in the cytosol (Fig. 6C).

**8.** The PomXY cluster acts as a barrier for the diffusion of PomZ dimers on the nucleoid (Fig. 6A). This might be due to the PomXY cluster acting as a PomZ sink and possibly also as a steric hindrance. Thus, PomZ dimers are considered to be strongly hindered from diffusing past the cluster in the model.

**9.** Based on quantitative western blot analysis (data not shown), *M. xanthus* cells contain ~200 PomX molecules, 850 PomY molecules and ~200 PomZ molecules. Based on fluorescence microscopy, ~45% of PomX, and ~15% of PomY are present in the PomXY cluster resulting in an estimated molecular weight of this complex of ~15 MDa. ~10% of PomZ molecules are in the PomXYZ complex at any one time and the remaining PomZ is bound to the nucleoid on either side of the cluster.

**Description of the computational model**

Here, we describe the details of our computational model. Fig. S7A illustrates our model with an emphasis on the numerical implementation. We model the nucleoid and the cluster as one-dimensional lattices of length $L$ and $L_{cluster}$, respectively (the lattice spacing we denote by $a$, resulting in a total of $M$ lattice sites) with reflecting boundaries at both ends of the nucleoid and the cluster. We assume that PomZ dimers can attach to every site on the nucleoid with the same probability. After ATP hydrolysis and detachment from the nucleoid and the cluster, PomZ molecules undergo nucleotide exchange and dimerization before they can reattach to the nucleoid. This delay between detachment and reattachment to the nucleoid together with fast diffusion of PomZ in the cytosol justifies our assumption of a constant attachment rate $k_{on}$ along the nucleoid. Diffusion of PomZ dimers on the nucleoid is modelled on the one-dimensional lattice as hopping between neighbouring lattice sites. The rate with which a PomZ dimer at site $n$ hops to the left or right (site $n-1$ or $n+1$) is set to $k_{hop}^0 = D/a^2$, because this results in a diffusive process with a diffusion constant $D$ in the continuum limit. Since the time the cluster takes to move to midcell is only about ¼ of the doubling time of the cell, we assume that the number of PomZ molecules in the cell is constant over the course of the simulation. The constant number of PomZ molecules in the cell leads to an attachment rate of PomZ dimers to the nucleoid that is proportional to the number of PomZ dimers in the cytosol $N_{cyto}$. Hence, the attachment rate of a PomZ dimer to one lattice site is given by $k_{on} \times N_{cyto}/M$, where $k_{on}$ is the attachment rate to the nucleoid. The simultaneous interactions of PomZ dimers with the PomXY cluster and the nucleoid are implemented similarly to a model introduced by Lansky et al. for crosslinker proteins bound between microtubules (Lansky et al., 2015). As discussed in the main text, we model the PomZ dimers as springs with a spring constant $k$. We take the rates for attachment of nucleoid-bound PomZ dimers to the cluster in a stretched state as the rate for attachment in the unstretched state, $k_a^0$, weighted by a Boltzmann factor:

$$k_a = k_a^0 \exp\left[-\frac{1}{2}\frac{k}{k_B T}\left(x_i^{cluster} - x_i^{nucleoid}\right)^2\right].$$

In the above equation, $k_B$ is the Boltzmann constant, $T$ the temperature and $x_i^{cluster}$, $x_i^{nucleoid}$ are the positions of the binding sites of the $i$-th PomZ dimer to the cluster and the nucleoid, respectively. Furthermore, a PomZ dimer bound to the nucleoid and the cluster is allowed to hop to the left or right lattice site with the rate (Lansky et al., 2015):

$$k_{hop} = k_{hop}^0 \exp\left[-\frac{1}{4}\frac{k}{k_B T}\left(\left(x_i^{cluster,to} - x_i^{nucleoid,to}\right)^2 - \left(x_i^{cluster,from} - x_i^{nucleoid,from}\right)^2\right)\right],$$

with $x_i^{cluster,from}$, $x_i^{nucleoid,from}$ and $x_i^{cluster,to}$, $x_i^{nucleoid,to}$ signifying the positions of the binding sites of the $i$-th PomZ dimer to the cluster and nucleoid before and after hopping, respectively. We assume that for the hopping events detailed balance holds. Hence, the ratio of the hopping rates from one site to a neighbouring one and back to the original site is weighted by a Boltzmann factor and the hopping rates are only determined up to a constant factor. We chose this factor such that the rate for hopping to a neighboring site and the rate for hopping back are the inverse of each other (Lansky et al., 2015). Note that we neglect the vertical distance between the nucleoid and the cluster when calculating the energy of a loaded spring. We introduce a rate $k_h$ for a PomZ dimer doubly bound to the cluster and the nucleoid to detach completely from the cluster and the nucleoid. This rate is assumed to be constant, i.e. it does not depend on the degree of stretching of the spring. Note that we neglect detachment of PomZ dimers, which are bound to the nucleoid but not to the cluster, because the ATP turnover rate of PomZ in contact with DNA is negligible compared to the turnover rate of PomZ in contact with DNA and PomX/PomY (Fig. 5C).

The cluster movement is described by a force balance equation

$$\gamma \frac{dx}{dt} = - \sum_{i=1}^{N} k\left(x_i^{cluster} - x_i^{nucleoid}\right),$$

stating that the frictional force exerted on the cluster, given by the friction coefficient $\gamma$ times the cluster velocity, balances the sum of all spring forces exerted on the cluster by the doubly bound PomZ dimers; the sum extends over all PomZ dimers bound to the cluster and the nucleoid (in total $N$).

In WT, approximately 10% of the 200 PomZ molecules are in the cluster. By contrast, PomZ[D90A]-mCherry, which cannot undergo ATP hydrolysis, is mostly in the cluster if expressed at native levels suggesting that in WT cells the concentration of PomZ dimers bound to the cluster is below its saturation limit. In cells overexpressing PomZ[D90A]-mCherry > 50-fold, the fusion protein is also bound to the nucleoid away from the cluster (Fig. 6B). We conclude that the PomXY cluster has a limited number of binding sites but at least binding sites for 100 PomZ dimers. To include the limited number of binding sites on the cluster and on the nucleoid in our model, we implemented a maximal density of binding sites on the cluster and the nucleoid, $c_{cluster}$ and $c_{nucleoid}$, respectively. When all binding sites of a lattice site are occupied, no further PomZ dimer can attach to this site from the cytosol or from the neighbouring lattice sites.

We implemented the model as a Gillespie algorithm (Gillespie, 1976; Gillespie, 1977; Lansky et al., 2015). Since the position of the cluster changes with time, the rates $k_a$ and $k_{hop}$ are time-dependent. For the parameters we consider here, the simulation results show that this time-dependence can be neglected and hence we assume that the rates are constant. Unless stated otherwise, we chose the initial PomZ distribution such that all PomZ dimers are in the cytosol and let the simulation run for at least 60 sec while keeping the cluster fixed to allow the PomZ dynamics to approach the steady state distribution. The initial position of the midpoint of the cluster, $p$, was at 0% or 10% of the nucleoid length.

**Simulation results**

At present, detailed experimental information about several of the molecular parameters of our computational model is lacking. Therefore, the focus of our mathematical analysis is mainly on the qualitative behaviour, and not a quantitative comparison between the results of the mathematical model and the experimental data. For some parameters it is possible to obtain an estimate from our experimental data and some parameters are estimated from the corresponding values from plasmid and chromosome segregation systems. The parameter estimates and sources are listed in Table S1. The detachment rate of PomZ dimers doubly bound to the cluster and the nucleoid can be estimated from the measurement of the ATP turnover rate, because we assume that ATP hydrolysis is the rate-limiting step. The stiffness of the springs mimicking the elasticity of the PomZ dimers and the chromosome is set to a value obtained from a model for an *in vitro* reconstituted ParAB*S* system (Hu et al., 2015).

For the diffusion constant of PomZ on the nucleoid, we chose a value similar to the experimentally determined values in plasmid and chromosome segregation systems. We chose the value of the friction coefficient from the related ParB/*parS* systems, relating the diffusion constant of the ParB/*parS* complex to its friction coefficient by the Stokes-Einstein equation, $D_{\text{cluster}} = k_B T / \gamma$. Increasing the friction of the PomXY cluster in the cytosol leads to a longer mean time for the cluster to reach midnucleoid. The value used in the simulations (Table S2) is the value that results in cluster trajectories where the cluster typically reaches midnucleoid in approximately the same time as observed experimentally. With the full set of parameters listed in Table S2, the model reproduces midnucleoid localization of the PomXYZ cluster with the same timing as observed experimentally in *M. xanthus* cells (Fig. 7C, S7BC). The simulation results are largely insensitive to changes of the attachment rate of PomZ dimers to the nucleoid and to changes of the binding rate of nucleoid-bound PomZ dimers to the cluster. We tested both parameters over several orders of magnitudes without noticeable deviations in the simulation results (Table S2). The density profile of PomZ dimers on the nucleoid is different for different cluster positions (Fig. 7D). Importantly, the concentration of PomZ is always highest, where the cluster is located (Fig. 7D). The total amount of PomZ in the Pom cluster increases, when the cluster moves from an off-centre position towards midcell. This is because the average distance that a PomZ dimer travels on the nucleoid until it reaches the cluster gets smaller the closer the cluster is located to midnucleoid. Importantly, if the cluster is positioned at an off-centre position, the density profile is asymmetric over the cluster and with the highest density of PomZ on the side of the cluster facing the more distant pole of the nucleoid. By contrast, if the cluster is at midnucleoid, PomZ has a symmetric distribution over the cluster (Fig. 7D). The same holds true for the density of PomZ dimers over the entire nucleoid: it is asymmetric if the cluster is located off-centre, with a higher density on the side with the longer cluster-to-nucleoid end distance, and becomes symmetric around the cluster, if the cluster is at midnucleoid (Fig. 7D). These predictions are in overall agreement with our experimental data in which we found that even though the patchy PomZ signal over the nucleoid is almost symmetrically distributed around the cluster, the patchy PomZ signal over the nucleoid is slightly but significantly higher in the case of cells with an off-centre cluster and with the highest intensity on the side of the cluster containing most of the nucleoid.

Our simulation results predict that the mechanism of midnucleoid positioning of the PomXYZ cluster would be disturbed if PomZ is overexpressed or if PomZ ATP hydrolysis is perturbed. If the number of PomZ molecules is increased in our simulations, the movement of the cluster is less biased towards midnucleoid (Fig. S7D). For twice the number of PomZ molecules, the clusters behave as for the WT case, for four-fold the PomZ molecule number the bias towards midcell is reduced, and for six-fold the PomZ molecule number the cluster is stalled at its initial position and moves only slightly. These predictions can be explained as follows: If the number of PomZ dimers is high, exclusion effects on the nucleoid and at the cluster become important and cluster motion is reduced in two ways: First, the flux difference of PomZ dimers on the nucleoid into the cluster decreases and PomZ dimers often cannot bind to the cluster because all binding sites are occupied. This leads to a reduced bias of the cluster movement towards midnucleoid. Second, crowding of PomZ dimers at the cluster impedes the mobility of PomZ dimers bound to the cluster and hence reduces cluster motion. These predictions agree with our experimental observations in the presence of extra PomZ, i.e. the PomXY cluster shows no directed movement towards midcell and little motion in cells overexpressing PomZ-mCherry > 50-fold (Fig. 4BC).

Cluster movement is also decreased for small ATP hydrolysis rates. ATP hydrolysis is necessary for the PomZ molecules to cycle between the cluster-bound and the cytosolic state. If the ATPase activity of PomZ is reduced, the number of PomZ dimers bound to the nucleoid is increased and ultimately, in the complete absence of ATP hydrolysis, all PomZ molecules become attached to the cluster because they cannot escape from the cluster once they are bound. Due to particle conservation in the cell, this leads to a decreased amount of PomZ molecules in the cytosol. Hence, the flux from the cytosol to the nucleoid becomes smaller and, therefore, the flux of nucleoid-bound PomZ into the cluster also decreases. This results in cluster trajectories that are less and less biased towards midcell for decreasing hydrolysis rates (Fig. S7E). Moreover, the higher PomZ dimer density at the cluster leads to similar crowding effects as in the case for PomZ overexpression. Again, the mobility of the PomZ dimers bound to the cluster is reduced and therefore the cluster movement is reduced. These predictions are in perfect agreement with our experimental observations with the ATP hydrolysis deficient variant PomZ$^{D90A}$. Importantly, if the ATP hydrolysis rate is decreased approximately 10-fold, the cluster typically undergoes a biased random motion towards midnucleoid, however, the simulations suggest that this translocation occurs more slowly. These predictions are in agreement with our experimental observations on PomX cluster dynamics in the absence of PomY (Fig. 4BC).

.

**Supplementary Experimental Procedures**

Cell growth and strains. DK1622 was used as WT *M. xanthus* strain and all strains are derivatives of DK1622 unless otherwise noted. In-frame deletions were generated as described (Shi et al., 2008). *M. xanthus* was grown at 32°C in 1% CTT medium (Hodgkin and Kaiser, 1977) or on 1.5% agar supplemented with 1% CTT and kanamycin (50µg/ml), oxytetracycline (10µg/ml) or gentamycin (10µg/ml) if appropriate. To induce the expression of genes from P$_{cuoA}$ in *M. xanthus* (Gómez-Santos et al., 2012), the growth medium was supplemented with copper sulfate as indicated in the text. Plasmids were integrated by site specific recombination into the Mx8 *attB* site or by homologous recombination at the native site, the *cuoA* locus or in the *mxan_18-19* intergenic region. All in-frame deletions and plasmid integrations were verified by PCR (Sambrook and Russell, 2001). *E. coli* was grown in LB or 2xYT medium (Sambrook and Russell, 2001). Plasmids were propagated in *E. coli* TOP10 (F- *mcr*A, Δ(*mrr-hsd*RMS-*mcr*BC), Φ80*lac*ZΔM15, Δ*lac*X74, *deo*R, *rec*A1). To induce expression of genes in *E. coli* for protein localization BL21 (DE3) (*fhuA2 lon ompT gal (λDE3) dcm ΔhsdS λDE3=λsBamHIo ΔEcoRI-B int::(lacI::PlacUV5::T7gene1) i21 Δnin*5) was used. Isopropyl-β-D-thiogalactopyranoside (IPTG) or arabinose was added as indicated.

Immunoblot analysis. Polyclonal α-PomX and α-PomY antibodies were raised by immunization of rabbits with the purified His$_6$-tagged proteins (Eurogentec). Immunoblot analysis was performed as described (Sambrook and Russell, 2001), using α-PomX, α-PomY, α-PomZ (Treuner-Lange et al., 2013), α-FtsZ (Treuner-Lange et al., 2013) or α-PilC (Bulyha et al., 2009) together with horseradish-conjugated goat anti-rabbit immunoglobulin G as recommended by the manufacturer (Sigma) as secondary antibody. For detection of mCherry-tagged proteins monoclonal rabbit α-mCherry antibodies were used as described by the manufacturer (BioVision) together with the peroxidase-conjugated goat α-rabbit immunoglobulin G secondary antibodies. Blots were developed using Luminata Forte chemiluminescence reagent (Millipore).

Purification of His$_6$-PomZ, PomY-His$_6$, PomX-His$_6$ and native FtsZ. Soluble His$_6$-PomZ was purified from *E. coli* as described (Treuner-Lange et al., 2013). To overexpress His$_6$-tagged PomY, plasmid pDS3 was propagated in *E. coli* ArcticExpress(DE3)RP cells (Agilent Technologies). Cells were grown at 30°C in LB medium to an OD$_{600}$ of 0.6–0.7. Cultures were pre-cooled and shifted to 18°C prior to induction of *pomY-His$_6$* expression with 1mM IPTG. Cells were incubated overnight at 18°C shaking at 230rpm. Cells were washed in lysis buffer 1 (50mM NaH$_2$PO$_4$; 300mM NaCl; 10mM imidazole; pH 8.0 (adjusted with NaOH)), and then lysed in 50ml lysis buffer 2 (lysis buffer 1; 0.1mM EDTA; 1mM β-mercaptoethanol; 100µg/ml PMSF; 1 x complete protease inhibitor (Roche Diagnostics GmbH); 10U/ml DNase 1; 0.1% Triton X-100) by 3 rounds of sonication for 5min with a Branson Sonifier (Duty cycle 4; output control 40%) (Heinemann) on ice. Cell debris was removed by centrifugation 4700rpm for 20 min at 4°C and additional filtration with a 0.45µm sterile filter (Millipore Merck). PomY-His$_6$ was purified with a 5ml HiTrap Chelating HP column, preloaded with NiSO$_4$ and equilibrated with lysis buffer 1. Proteins were eluted with elution buffer (lysis buffer 1 supplemented with 500mM imidazole). Elution fractions containing PomY-His$_6$ were loaded onto a HiLoad 16/600 Superdex 200 pg column that was equilibrated with dialysis buffer (50mM Hepes/NaOH pH 7.2; 50mM KCl; 0.1mM EDTA; 1mM β-mercaptoethanol; 10% (v/v) glycerol). Collected PomY-His$_6$ after gel filtration was pooled and concentrated with a 5ml HiTrap SP HP column, equilibrated with dialysis buffer. PomY-His$_6$ was eluted from the column with cation-exchange buffer (dialysis buffer with 2000mM KCl) using a 50mM to 2000mM KCl gradient. Elution fractions from the ion-exchange chromatography were dialyzed against dialysis buffer at 4°C. To purify PomX-His$_6$ plasmid pEMR3 was propagated in *E. coli* NiCo21 DE3 cells (NEB). Cells were grown in LB medium with 50µg/ml kanamycin at 30°C to an OD$_{600}$ of 0.6 – 0.7. Protein accumulation was induced with 0.5mM IPTG for 16 hrs at 18°C. Cells were washed in lysis buffer 1 and lysed in 50ml lysis buffer 2 without Triton X-100 by sonication as described before. Cell debris was removed by centrifugation at 20000g for 20 min a 4°C and PomX-His$_6$ was affinity purified with Protino® Ni-NTA resin (Macherey-Nagel) from batch, equilibrated in Lysis buffer 1. PomX-His$_6$ was eluted from the resin by washing 1 x with 5ml elution buffer 1 (lysis buffer 1 with 50mM imidazole) and 3 x with 5ml elution buffer 2 (lysis buffer 1; 250mM imidazole). Purified PomX-His$_6$ was dialyzed against dialysis buffer. To purify native FtsZ plasmid pKA70 was propagated in Rosetta2(DE3) cells (Novagen). Cells were grown in 2xYT medium with 50µg/ml ampicillin at 37°C to an OD$_{600}$ of 0.6 – 0.7. Expression of *ftsZ* was induced with 0.5mM IPTG for 3 hrs at 37°C. Cells were washed in FtsZ lysis buffer 1 (50mM Tris-HCl pH 7.9; 50mM KCl; 1mM EDTA; 1mM β-mercaptoethanol; 10% (v/v) glycerol) and lysed in FtsZ lysis buffer 2 (lysis buffer 1, 1 x complete protease inhibitor (Roche Diagnostics GmbH); 10U/ml DNase 1) by sonication as described above. Cell debris was removed by centrifugation at 20000g for 20 min at 4°C and the soluble fraction was applied to ammonium sulfate precipitated by adding ammonium sulfate in small steps to 33% of saturation at 4°C to precipitate native FtsZ. Precipitate was dissolved in lysis buffer 1 and loaded onto a 5ml HiTrap Q HP column, equilibrated in lysis buffer 1. Proteins were eluted with FtsZ elution buffer (FtsZ lysis buffer 1; 1000mM KCl). Fractions containing FtsZ were pooled and diluted 1:5 with dilution buffer (lysis buffer 1 without KCl) and loaded onto a 5ml HiTrap Q HP column, equilibrated with lysis buffer 1. Proteins were eluted again with FtsZ elution buffer as described before along a short gradient of 4 column volumes.

Protein sedimentation assay. Before sedimentation assays, a clearing spin was performed for proteins to be analyzed at 20,000g for 10min at 4°C. Purified proteins at a final concentration of 3µM in a total volume of 25µl were mixed and incubated at for 2-10min at 32°C in buffer (50mM Hepes/NaOH, pH 7.2, 50mM KCl, 1mM ß-mercaptoethanol, 10mM MgCl$_2$). Subsequently, samples were separated into soluble and insoluble fractions by centrifugation (160,000g, 60 min, 25°C). Insoluble and soluble fractions were separated. Equivalent volumes of soluble and the insoluble fractions were separated by SDS-PAGE and stained with Instant Blue™ (expedeon) for 10 min.

ATPase assay. A colorimetric ATPase assay was performed as described (Treuner-Lange et al., 2013). Briefly, His$_6$-PomZ alone or with PomY-His$_6$, PomX-His$_6$ or both at final concentrations of 2µM were mixed in a 96-well plate (Greiner Bio-One) in triplicate in buffer A (50mM Hepes/NaOH pH 7.2, 50mM KCl, 10mM MgCl$_2$, 1mM β-mercaptoethanol, 1mM ATP). Reactions were incubated for 30 min at 37°C. Reactions were mixed with 250µl Malachite-green reagent (Sigma-Aldrich) incubated for 5 min and stopped with 50µl of 34% (v/v) citric acid. After 15 min, the developed color was measured with an Infinite M200 Pro plate-reader at 660nm (Tecan). If nonspecific DNA was added, pUC18 plasmid DNA in 50mM Hepes/NaOH pH 7.2; 50mM KCl; 0.1mM EDTA; 1mM β-mercaptoethanol; 10% (v/v) glycerol was used at a final concentration of 5nM.

Negative stain transmission electron microscopy. For fixation and negative stain of protein samples, 10µl of a protein sample of interest (concentration before applying: PomX-His$_6$ and PomY-His$_6$, 3µM and 3µM) were applied on one side of the EM grid (Plano) and incubated for 1min at room temperature. Liquid was blotted through the grid by applying the unused side of the grid on Whatman paper. The grid was washed twice with double-distilled H$_2$O and once with a 1% uranyl acetate solution with the same technique. Then uranyl acetate was applied on the grid for 20 sec and dried by blotting the liquid through the grid with a Whatman paper. In case of FtsZ, 3µM protein was pre-incubated at room temperature for 2 min before GTP was added at a final concentration of 2.5mM. After additional incubation for 10 min, protein was applied to the grid and grids were handled as described above. Finished grids were stored in a grid holder for several months at room temperature. Electron microscopy was done with a CM120 elcetron microscope (FEI) at 120kV.

Right angle light scattering. Right angle light scattering was performed with 10µM FtsZ at pH 6.5 (50mM MES/NaOH, 50mM KCl, 10mM MgCl$_2$, 1mM β-mercaptoethanol) on a temperature-controlled ISS PC1 spectrofluorometer with a cooled photomultiplier with excitation and emission wavelength set to 350nm. FtsZ was preincubated for 2 min at 8°C and experiment was subsequently started. After 100 sec GTP at a final concentration of 2.5mM was added to initiate filament formation. Light scattering was followed for 900 sec. FtsZ without addition of GTP served as negative control.

GAL4-based yeast two hybrid assay. Yeast two hybrid assays were performed as described by the manufacturer (Clontech). Briefly, genes of interest were fused to the GAL4-AD fragment (activation domain) or the GAL4-BD fragment (DNA-binding domain). Plasmids were co-transformed into yeast strain AH109. Transformants were selected on SD/-Leu/-Trp agar for inheritance of both plasmids. Four independent clones were resuspended in SD/-Leu/-Trp medium and grown for 3 doubling times at 30°C. OD was adjusted to 0.5 and 3µl cells were placed on SD/-Leu/-Trp/-His (medium stringency) selective agar. Plates were incubated at 30°C for 120 hrs. Growth on medium stringency selective agar was classified as a positive interaction. Additionally each plasmid containing a gene of interest was co-transformed with an empty vector only expressing the GAL4-AD or BD fragment, respectively. In an experiment, all strains were spotted with all the controls on the same selective agar plates. The data shown is a representative agar plate of two independent transformation experiments.

Fluorescence microscopy. Fluorescence recovery after photobleaching (FRAP) experiments were performed with a temperature controlled Nikon Ti-E microscope with Perfect Focus System and a CFI PL APO 100x/1.45 Lambda oil objective at 32°C with a Hamamatsu Orca Flash 4.0 camera using NIS Elements AR 2.30 software (Nikon). For photobleaching a 651nm laser beam was focused on the on the central part of the image plane. After acquisition of an initial pre-bleach picture, cells of interest were bleached using a single 5 × 5 pixel circular shaped region. Photobleaching of SA3131 was performed with 1 laser pulse with 5% laser power and a dwelling time of 500 µsec. For strains SA7011 and SA4799, FRAP was performed with 10 consecutive laser pulses with 10% laser power using the same dwelling time. Images were recorded every 1 sec in case of SA3131 and every 300 msec in case of SA7011 and SA4799. For every image, total integrated cellular fluorescence in a region of interest (ROI) within the outline of the cell was measured together with total integrated background fluorescence of a ROI of the same size placed outside of the cell. Additionally, fluorescence intensity was measured in the bleached region together with background fluorescence or a ROI of same size placed on the background. After background correction, corrected fluorescence intensity of the bleached area (or area of interest) was divided by total corrected cellular fluorescence, which in term corrects for bleaching effects during picture acquisition. This relative fluorescence was correlated to the initial fluorescence in the bleached area (or area of interest). The mean relative fluorescence of several cells was plotted as function of time [sec]. To determine the recovery rate for the tested fluorescent protein ($t_1$), the plotted data was fitted to a single exponential equation ($y = y_0 + A * e^{-x/t}$) using Origin8.0. Half-maximal recovery ($t_{1/2}$) was calculated from the recovery rate ($t_1$) by $t_{1/2} = \ln(2) * t_1$.

For PomZ asymmetry analysis, PomZ-mCherry in SA3131 was imaged with an acquisition time of 1 sec. To calculate the asymmetry index the integrated and background-corrected patchy PomZ-mCherry signal on either side of a cluster was divided by the fluorescent area on either side of the cluster, giving the intensity values $I_{left}$ and $I_{right}$, left and right of the PomZ-mCherry cluster (Fig. S3B). The asymmetry index was calculated as $(I_{left}-I_{right}) / (I_{left}+I_{right})$ independently for off-centre and midcell clusters. The normalized asymmetry value for the nucleoid on either side of the cluster was calculated similarly using Pico green stained nucleoids (Harms et al., 2013) in cells of SA3131.

Plasmid construction. All DNA fragments generated by PCR were verified by sequencing.

Plasmids pDS1, pDS12, pDS16, pMAT12, pAH27: For pDS1 up- ("KA-371/Mxan_0634-5") and downstream fragment ("KA-373/KA-374") were amplified from genomic *M. xanthus* DNA and digested EcoRI+XbaI and XbaI+HindIII, respectively. Fragments were cloned separately into pBJ114 and sequenced. For pAH27 up- ("KA-224/Mxan_0636-1") and downstream fragment ("Mxan_0636-2/Mxan_0636-3") were amplified from genomic *M. xanthus* DNA and digested EcoRI+XbaI and XbaI+HindIII, respectively. Fragments were cloned separately into pBJ114 and sequenced. pMAT12 is a derivative of pAH27. Upstream fragment ("KA-200/Mxan_0635-3") was amplified from genomic DNA, digested with EcoRI+XbaI and cloned into pAH27 that was digested EcoRI+XbaI before. pDS12 is a derivative of pDS1. For construction of pDS12, pAH27 was digested XbaI+HindIII and downstream fragment was cloned into pDS1 that was digested with the same enzymes before. pDS16 is a derivative of pDS1 in which the downstream fragment (XbaI+HindIII) was replaced with another downstream fragment ("Mxan_0635-1/Mxan_0635-2") that was digested in the same way.

Plasmid pDS3 and pEMR3: For pDS3 *pomY* was amplified with "Mxan_0634-11" and "Mxan_0634-12" from genomic DNA and cloned into pET24b+ (EcoRI and HindIII). For pEMR3, *pomX* was amplified with primer "NdeI-pomX fwd" and "pomX c-term His rev" and cloned into pET24b+ using NdeI and HindIII.

Plasmids pDS81, pDS82: For both plasmids *pomY* was amplified from genomic *M. xanthus* DNA using the primers "Mxan_0634 fwd EcoRI" and "Mxan_0634 rev stop BglII" and cloned into pGAD424 and pGBT9.

Plasmids pDS83, pDS84: For pDS83 *pomZ* was amplified from genomic *M. xanthus* DNA using the primers "Mxan_0635 fwd SalI linker" and "Mxan_0635 rev stop BglII" and cloned into

pGAD424. For pDS84 *pomZ* was amplified using the primers "Mxan_0635 fwd BamHI linker" and "Mxan_0635 rev stop PstI" To be translated in the correct frame *pomZ* contains one additional amino acid at the beginning which additionally serves as a part of the linker for both GAL4-AD and GAL4 DNA-BD fragment fusions.

Plasmids pDS85, pDS86: For both plasmids *pomX* was amplified from genomic *M. xanthus* DNA using the primers "Mxan_0636 fwd EcoRI" and "Mxan_0636 rev stop BamHI" and cloned into pGAD424 and pGBT9.

Plasmids pDS87, pDS88: For both plasmids *ftsZ* was amplified from genomic *M. xanthus* DNA using the primers "Mxan_5597 fwd EcoRI" and "Mxan_5597 rev stop BamHI" and cloned into pGAD424 and pGBT9.

Plasmids pDS7, pDS8, pDS18 and pDS19: For pDS8 *pomY* with its native promoter was amplified from genomic DNA with primers "KA-371" and "Mxan_0634-4" and cloned with EcoRI and BglII into pKA28 (Treuner-Lange et al., 2013). For pDS7 *pomY-mCherry* was amplified from pDS8 with "Mxan_0634-6" and "mCherry stop rev HindIII" and cloned into pSW105 with XbaI and HindIII. For pDS18 *eyfp* was amplified from pAH7 with "KA-396" and "KA-397" and cloned with BamHI and HindIII into pSWU30. *pomY* was amplified with "KA-371" and "Mxan_0634-4" from genomic DNA and cloned into the resulting plasmid with EcoRI and BglII. P$_{pilA}$*pomY-mCherry* was excised from pDS7 with EcoRI and HindIII and ligated into pSWU30 resulting in pDS19.

Plasmids pKA46, pAH35, pAH52, pAH53 and pAH96: For pKA46 the native *pomX* promoter was amplified with "KA-382" and "KA-383" from genomic DNA and cloned with EcoRI and KpnI into pSWU30. *mCherry* was amplified with "KA-302" and "KA-303" from pKA28 and cloned into the same plasmid using KpnI and BamHI. For the third fragment *pomX* was amplified with "KA-384" and "KA-348" from genomic DNA and cloned into the vector with BamHI and HindIII. For pAH35 *mCherry-pomX* was amplified from pKA46 with primers "mCherry fwd XbaI" and "KA-348" and cloned into pSW105 using XbaI and HindIII as restriction sites. The same strategy was used for pAH52 but *mCherry-pomX* was cloned into pMAT11 with XbaI and HindIII. Plasmid pAH53 is a derivative of pAH35 in which the *pilA* promoter is replaced by the native *pomZ* promoter that was amplified with "KA-200" and "pomZ prom rev XbaI" and cloned into pAH35 with EcoRI and XbaI. For pAH96 P$_{pomZ}$*mCherry-pomX* was excised from pAH53 with EcoRI and HindIII and cloned into pSWU30.

Plasmids pDS21 and pDS22: For pDS21 *pomY-mCherry* was excised from pDS7 with XbaI and HindIII and cloned into pMAT11. For pDS22 *pomY-eyfp* was amplified from pDS18 with primers "KA-397" and "Mxan_0634-6" and ligated XbaI-HindIII into pMAT11.

Plasmids pEB16, pKA55 and pAH100: All three plasmids are derivatives of pKA28 and constructed by site directed mutagenesis using the Quickchange II XL site directed mutagenesis kit (Stratagene) as described by the manufacturer's instructions. For pEB16 point mutation for *pomZ*$^{G62V}$ was introduced with primers "KA-413" and "KA-414". For pKA55 point mutation for *pomZ*$^{K66Q}$ was introduced with primers "KA-417" and "KA-418". For pAH100 point mutation for *pomZ*$^{K268E}$ was introduced with primers "AH-91" and "AH-92".

Plasmid pKA45 and pEB13: For pKA45 *pomZ-mCherry* was amplified from pKA28 with "KA-288" and "KA-293" and cloned XbaI-HindIII into pSW105. pEB13 was constructed in the same way by using pKA43 as template for PCR fragment.

Plasmid pAH83: A fragment downstream of *ssb* (*MXAN_1071*) was amplified with primers "EB-7" and "EB-8" from chromosomal DNA and cloned BamHI and EcoRI into pBJ114. Then *eyfp* was amplified from pAH7 with primers "EB-11" and "EB-12" and cloned into the same plasmid with BamHI and XbaI. Finally, a fragment upstream of *ssb* was amplified with primers "EB-5" and "EB-6" and also cloned into the same plasmid using XbaI and HindIII, resulting in pAH83.

Plasmid pDS74, pDS75 and pDS80: For pDS74 $P_{pilA}pomZ$ was excised from pKA19 with EcoRI and HindIII and cloned into pSWU30 digested with the same enzymes. pDS75 is a derivative of pDS74. For pDS75 the Mx8 *attB* locus was excised from pDS74 using BsrDI and BlpI. This was replaced by the *MXAN18-19* intergenic region amplified from pMR3691 using the primers "Mxan18-19 fwd BsrDI" and "Mxan18-19 rev BlpI". To create pDS80, $P_{nat}pomZ^{D90A}$ together with a short part of the vector backbone was excised from pKA43 with NdeI and HindIII. This fragment was cloned into pDS75 to replace $P_{pilA}$ *pomZ*, which was digested with the same enzymes before.

Plasmid pDS37, pDS43 and pDS46: pDS37, pDS43 and pDS46 are derivatives of pRSFDuet-1. To construct pDS37 *mCherry-pomX* was amplified from pAH53 with "mCherry BspHI fwd" and "Mxan_0636-3 HindIII rev stop" and cloned into pRSFDuet-1 multiple cloning site 1 (MCS1) with BspHI and HindIII. For pDS43 *pomZ-mCherry* was amplified from pKA28 with primers "Mxan_0635 BspHI fwd" and "KA-478" and cloned into pRSFDuet-1 in the same way. For pDS46 *pomY-eyfp* was amplified from pDS18 using "Mxan_0634-18" and "Mxan_0634-15" and cloned into pRSFDuet-1 as described before.

Plasmid pDS68: To construct pDS68, *pomZ-mCherry* was amplified from pKA28 with "Mxan_0635 start NdeI" and "mCherry stop PacI" and cloned into MCS2 of pDS46 using the indicated restriction enzymes.

Plasmids pDS38 and pDS45: For pDS38 *pomX* was amplified from genomic DNA using the primers "Mxan_0636 BspHI fwd" and "Mxan_0636-3 HindIII rev stop" and cloned into pBAD24 using the indicated restriction sites. To construct pDS45, *pomY-eyfp* was amplified from pDS18 with primers "Mxan_0634-18" and "Mxan_0634-15" and ligated into pBAD24 using BspHI and HindIII restriction sites.

Plasmid pDoB12: For pDoB12 *ftsK* was amplified with the primers "*ftsK* start XbaI" and "*ftsK* nostop BamHI rev" from genomic DNA and cloned with the indicated enzymes into pKA45.

Plasmid pMAT112: For pMAT112 the *aadA* gene was amplified with "pIJ778 SacI down" and "pIJ778 BglII Pst up" from pIJ778 and ligated into pMAT76 that was amplified with "pMAT76 BglII" and "pMAT76 SacI". For pMAT76, the multiple cloning site of pMR3691 was changed. pMR3691 was opened with NdeI and KpnI and ligated with an annealed primer double strand consisting of "Cla-Sca linker+" and "Cla-Sca linker –", resulting in pMAT56. This vector was opened with ClaI and KpnI and ligated with *tetR-eyfp* that was amplified from pMAT6 with "TetR-YFP-ClaI" and TetR-YFP-KpnI.

**Legends to Supplementary Figures**

Figure S1: *pomX*, *pomY* and *pomZ* are encoded in a gene cluster present in other myxobacteria
A. Conservation of the *pomXYZ* gene locus in Myxobacteria. Transcription direction is indicated by the orientation of arrows. PomX, PomY and PomZ homologs were identified with reciprocal BLASTP analysis (Huntley et al., 2011). % of similarity/identity between homologs is indicated by numbers in the arrows and were calculated using EMBOSS Needle software (pairwise sequence alignment) (Li et al., 2015).
B. *pomX*, *pomY* and *pomZ* gene cluster in *M. xanthus*. Start and stop codons are indicated. Domain structures were predicted by SMART analysis (Letunic et al., 2015). Transcription direction is indicated by the arrows.
C. Quantification of average cell length, constriction frequency and anucleate cells of the indicated strains. The strains expressing mCherry PomX$^{OE}$ and PomY-mCherry$^{OE}$ overexpress the two proteins (see also S1E). Strains used from top to bottom: DK1622, SA4223, SA4229, SA4252, SA4703, SA4713, SA4712, SA3108, SA4743, SA6130, SA4254, SA4722. (n>200 cells each strain).
D. Cell divisions in Δ*pomX* and Δ*pomY* mutants. Cell division position was analyzed in DAPI-stained cells of the indicated genotypes. White arrows indicate cell division constrictions. Scale bar 2μm.
E. Accumulation of PomX, PomY and their fluorescent fusions in strains of indicated genotypes. Equal amounts of protein were loaded per lane. Western blots were probed with α-PomX and α-PomY antibodies. The strains labelled mCherry PomX$^{OE}$ and PomY-mCherry$^{OE}$ overexpress the two proteins. Strains used in upper blow from left to right: DK1622, SA4223, SA4229, SA4252, and in the lower blot: DK1622, SA4703, SA4713, SA4712.
F. Immunoblot analysis of PomY and PomY-mCherry accumulation in cells of indicated genotypes grown in media supplemented with indicated concentrations of CuSO$_4$. From left to right: DK1622 (WT), SA4703 (Δ*pomY*), SA4734 (Δ*pomY*/P$_{cuoA}$*pomY-mCherry*). Equal amounts of protein were loaded per lane and the blots were probed with specific α-PomY antibodies as indicated.
G. Immunoblot analysis of PomZ, PomZ-mCherry and PomZ$^{D90A}$-mCherry accumulation after overexpression (OE). Equal amounts of protein were loaded per lane and the blots were probed with specific α-PomZ antibodies. Black arrow indicate PomZ and PomZ-mCherry, grey arrows indicate PomZ-mCherry degradation products. Strains used from left to right: DK1622, SA3108, SA3147, SA5006.

Figure S2: Δ*pomX* and Δ*pomY* mutants are unaffected in nucleoid localization and numbers
A. Quantification of chromosome number and number of origins of replication in WT, Δ*pomX* and Δ*pomY* cells. For nucleoid number analysis cells were stained with DAPI and in the case of WT additionally incubated with cephalexin for 8 and 12 hrs before visualization by DIC and fluorescence microscopy. The number of nucleoids was tracked and plotted as function of cell length (n>200 cells each strain). Strains used for quantification of nucleoids: DK1622, SA4223, SA4703. ParB-eYFP foci analyzed in SA4202 (*parB*$^+$/P$_{nat}$*parB-eyfp*), SA4219 (Δ*pomX*, *parB*$^+$/P$_{nat}$*parB-eyfp*), SA4709 (Δ*pomY*, *parB*$^+$/P$_{nat}$*parB-eyfp*).
B, C. Replication and chromosome segregation is unaffected by lack of PomX and PomY. The localization of nucleoid midpoints and ParB-eYFP were analyzed from the same cells as in A and plotted as a function of % of cell length. Histograms show the localization of midnucleoid and ParB-eYFP foci in cells with 1 (red), 2 (blue) and 4 (green) chromosomes or ParB-eYFP foci. Numbers display the mean position ± SD in percent of cell length. Scale bar, 2µm.

Figure S3. PomX and PomY localize at midcell before the end of replication.
A. PomX and PomY colocalize with cell division constrictions. White arrows indicate constrictions in SA4229 (upper panel) and SA4713 (lower panel). Scale bar, 2µm.
B. PomZ-mCherry localization in DAPI-stained cells. Linescans display fluorescence intensity of PomZ-mCherry signal (red) and DAPI signal (blue) along the long cell axes of the indicated cells. Scale bar, 2µm. Strain used: SA3131. Schematic PomZ-mCherry localization (bottom) illustrates the values $I_{left}$ and $I_{right}$ used for calculation of asymmetry values.
C. PomX and PomY localize to midcell before completion of replication. Δ*pomX* and Δ*pomY* cells expressing mCherry-PomX and PomY-mCherry together with Ssb-eYFP were DAPI stained and analyzed. Numbers display percentages with cells of that localization pattern (n=200 for each strain). Scale bar, 2µm. White arrows display off-centre clusters and orange arrows display midcell clusters of mCherry-PomX and PomY-mCherry. Strains used from top to bottom: SA7042, SA7043.
D. PomZ recruits PomX and PomY clusters to the nucleoid. Linescans as in B with DAPI (blue) and mCherry-PomX or PomY-mCherry (red). White arrows indicate clusters that do not colocalize with the nucleoid. The table lists % of cells with the localization on the left in the presence or absence of PomZ (n>200). Scale bar, 2µm. Strains used: SA4712, SA4720, SA4252, SA5821.
E. PomX and PomY colocalize in the absence of PomZ. White arrow indicates off-centre cluster of PomY-YFP and mCherry-PomX. Scale bar, 2µm. PomY-YFP was expressed in presence of 150µM $Cu^{2+}$. Strain used: SA5839.

Figure S4: PomX and PomY localize at midcell in the absence of FtsZ
A. Localization of mCherry-PomX and PomY-mCherry in FtsZ-depleted cells. Cells of strain SA5809 (left) and SA4718 (right) were treated as described in Fig. 3B and images acquired before and 12 hrs after removal of $Cu^{2+}$ from the growth media. White arrows indicate off-centre clusters and orange arrows indicate midcell clusters of mCherry-PomX and PomY-mCherry, respectively. Scale bar, 5µm.
B. Localization of PomY-mCherry and mCherry-PomX in cephalexin treated cells. Cells of strain SA4712 and SA4252 were treated with cephalexin for 12 hrs, stained with DAPI and analyzed by fluorescence microcopy. Mean cell length ± SD as well as constriction frequency were calculated (n>100 cells per time point and strain) together with PomY-mCherry and mCherry-PomX localization pattern at each time point. Pattern abundance is displayed in the histograms as % of total cells analyzed.
C. FtsK cluster formation depends on FtsZ. FtsK-mCherry localization was followed during an FtsZ-depletion experiment as in A at indicated time points using strain SA4169. Upper panel, FtsZ level during the depletion experiment. Middle panel, accumulation of the loading control PilC in the same cells. Lower panel, accumulation of FtsK-mCherry in the same cells. Cells were washed twice with copper-free medium and transferred to copper-free medium at t= 0 hrs. For each time point n>150 cells were analyzed to quantify average cell length ± SD, constriction frequency and pattern of FtsK-mCherry localization. Pattern abundance is displayed in the histograms as % of total cells analyzed. Representative cells are shown in the panels on the right. Scale bar, 2µm.

D. SDS-PAGE analysis of purified proteins used in this study. Proteins were applied to SDS-PAGE on a 10% SDS-gel and stained with Instant Blue™. Molecular size markers are shown on the left. The calculated molecular mass of PomX-His$_6$ is 45.4kDa, of PomY-His$_6$ 72.3kDa, of His$_6$-PomZ 37.7kDa and of FtsZ 44.7kDa. Note that in SDS-PAGE PomX-His$_6$ has a mobility larger than the monomer.
E. FtsZ forms GTP-dependent filaments in right angle light scattering. Experiments were performed with 10µM FtsZ at 8°C at pH 6.5. GTP at a final concentration of 2.5mM was added at 100 sec as indicated by the black arrow. The graphs show mean values from two independent experiments.
F. Negative stain electron microscopy of FtsZ filaments. 3µM FtsZ at pH 6.5 were applied to a grid and negatively stained after incubation with or without 2.5mM GTP for 10 min at room temperature. Scale bar, 100nm.

Figure S5. PomX and PomY are dynamically localized.
A. PomX and PomY colocalize during translocation to midcell and at midcell. Cells of strain SA7041 were recorded for 4 hrs on an agar pad containing 0.25% CTT growth medium at 32°C. Images were acquired every 15 min. PomY-eYFP was expressed in presence of 150µM $Cu^{2+}$. The white stippled line indicates a cell division event that is marked by a black arrow in the schematic. Scale bar, 2µm. Right, schematic illustrate cluster localization in the cell on the left. The black arrow indicates the cell division event.
B. PomX and PomY are occasionally asymmetrically distributed to daughter cells during cell division. Cells expressing mCherry-PomX (SA4797) or PomY-mCherry (SA7000) were followed for 4 hrs on an agarose pad containing 0.25% CTT growth medium at 32°C. Images were recorded every 15 min. The depicted cells are representative for an asymmetric distribution of mCherry-PomX and PomY-mCherry during cell division. White stippled lines indicate cell divisions. White arrows indicate mCherry-PomX and PomY-mCherry in an emerging cluster. Scale bar, 2µm.

Figure S6. PomZ ATPase activity and DNA-binding are important for PomZ function and localization
A. PomX and PomY cosediment *in vitro*. Instant Blue™-stained SDS-PAGE of sedimentation-reactions of 3µM of PomX-His$_6$ and PomY-His$_6$ alone and in combination. Protein content of supernatant (S) and pellet fraction (P) was separated after high speed centrifugation. Molecular size markers are shown on the left.
B. Schematic of canonical ParA ATPase cycle. Amino acid substitutions refer to the numbering of PomZ residues (Cf. H).
C. Subcellular localization of PomZ-mCherry variants in *E. coli*. *E. coli* BL21 DE3 was treated and analyzed as described in Fig. 5A. Scale bar, 2µm.
D. Subcellular localization of PomZ-mCherry variants in *M. xanthus*. Δ*pomZ* cells expressing PomZ, PomZ$^{K66Q}$, PomZ$^{G62V}$, PomZ$^{K268E}$ or PomZ$^{D90A}$ fused to mCherry were DAPI stained. White and orange arrows indicate off-centre and midcell clusters. Scale bar, 2µm. Strains used top to bottom: SA3131, SA5001, SA5000, SA5837, SA3146.
E. Immunoblot analysis of PomZ and PomZ-mCherry accumulation and its variants in cells of indicated genotypes. Strains used left to right: DK1622, SA3108, SA3131, SA5001, SA5000, SA5837, SA3146. Equal amounts of protein were loaded per lane and the blots were probed with α-PomZ antibodies.

F. PomZ[D90A] colocalizes with PomY clusters on the nucleoid. White arrow indicates colocalizing PomY-YFP and PomZ[D90A]-mCherry cluster in a cell for which the linescan is shown on the right; linescan was done as in Fig. 2A with DAPI (blue), PomZ-mCherry (red) and PomY-YFP (yellow). Scale bar, 2µm. PomY-YFP was expressed with 150µM $Cu^{2+}$ (Cf. Fig. S1F). Strain used: SA4758.

G. PomZ[D90A] cluster formation depends on PomX. Linescan is as in Fig. 2A with DAPI (blue) and PomZ[D90A]-mCherry (red). Scale bar, 2µm. Strain used: SA7014.

H. Alignment of PomZ with other ParA ATPases. PomZ was aligned with with other ParA ATPases using the MAFFT algorithm. Orange, brown, red and purple boxes indicate residues Gly62, Lys66, Asp90 and Lys268, respectively in PomZ. Note that only the parts of the alignment that include these four residues are shown. Protein sequences used for the alignment: *B. subtilis* Soj (gi|586852); *V. cholerae* ParA1 (gi|15642766); *V. cholera* ParA2 (gi|15601863); *P. aeruginosa* ParA (gi|15600756); *M. xanthus* ParA (gi|108467427); *C. crescentus* ParA (gi|239977514); *M. tuberculosis* ParA (gi|923109897); *C. glutamicum* ParA (gi|41223089); *S. coelicolor* ParA (gi|75489208); *R. sphaeroides* PpfA (gi|332276184); *M. xanthus* PomZ (gi|108460931).

Figure S7. A PomZ flux-based model for midcell positioning of the PomXYZ complex

A. Numerical implementation of the model. The nucleoid and the cluster are modeled as one-dimensional lattices with reflecting boundaries. A PomZ dimer in the cytosol can attach to a single lattice site of the nucleoid with rate $k_{on}/M$ (M is the total number of lattice sites of the nucleoid). PomZ dimers bound to the nucleoid can hop to a neighboring lattice site with rate $k_{hop}^0$, if the maximal number of binding sites is not reached for this lattice site (we chose the parameters such that only one dimer can bind per lattice site, Table S2). Due to thermal fluctuations nucleoid-associated PomZ dimers can spontaneously switch from a stretched to an unstretched conformation (indicated by transparent PomZ dimers). PomZ dimers on the nucleoid can bind to the cluster with a rate $k_a^0$, which decreases according to a Boltzmann factor if the PomZ dimer attaches to the cluster in a stretched state (rate $k_a$, the decreasing intensity of the arrows indicates a decrease in the attachment rate the further the PomZ dimer has to be stretched in order to attach to this site). Moreover, doubly-bound PomZ dimers can hop on the cluster and the nucleoid with rate $k_{hop}$ and hydrolyze ATP with rate $k_h$ and subsequently detach from the nucleoid and the cluster as two ADP-bound PomZ monomers into the cytosol. The position of the midpoint of the cluster on the nucleoid is denoted by $x$.

B. Simulated trajectories of the PomXY cluster for the parameters summarized in Table S2. The position of the midpoint of the cluster on the nucleoid over time is shown for 100 runs of the stochastic simulation in black. The data was divided with respect to time into 90 time intervals of the same size and the cluster positions were averaged per time interval (shown in red). The horizontal dashed grey line indicates midnucleoid.

C. Histogram of the time the cluster needs to reach midcell (same data as in B). Time is recorded until the midpoint of the cluster first reaches midnucleoid. Note that the average time a cluster needs to reach midcell is not equivalent to the time, when the average trajectory of all cluster movements reaches midcell. In red the mean ± SD of the distribution is shown.

D. Simulated trajectories of the PomXYZ cluster as described in B for 2, 4 and 6 times the number of PomZ dimers.

E. Simulated trajectories of the PomXYZ cluster as described in B, but for different ATP hydrolysis rates of PomZ dimers interacting with the PomXY cluster ($k_h = 0.1 \sec^{-1}, 0.001 \sec^{-1}, 0 \sec^{-1}$).

F. Effect of hydroxyurea (HU) on PomY-mCherry localization. Exponentially growing cells of strain SA4712 (Δ*pomY*/P*{pilA}pomY-mCherry*) were supplemented with 50mM HU for 16 hrs. Cells were treated with DAPI and imaged before treatment and then every 8 hrs after addition of HU. For each time point >100 cells were analyzed. To correlate PomY-mCherry cluster localization with chromosome length, the long axes of the DAPI stained nucleoid was measured.

G. Images of PomY-mCherry in HU-treated and untreated cells. Nucleoids were stained with DAPI. White arrows indicate off-centre clusters of PomY-mCherry and orange arrows indicate midcell clusters. Cells marked with white asterisk (*) were used for linescans (right). Linescans show fluorescence intensity of DAPI stained nucleoids (blue) and PomY-mCherry signals (red) along the long axes of the cell.

**Table S1**. Parameters of relevance for the Pom system from experiments and/or literature on related systems

| Parameter | Value | Source/Comment |
|---|---|---|
| Number of PomZ molecules per cell | 200 | Western plot analysis (data not shown) |
| Length of *M. xanthus* nucleoid | $4.8 \pm 1.3$ μm | Measured |
| Diameter of Pom cluster (along long and short axis) | $0.71 \pm 0.15$ μm<br>$0.63 \pm 0.09$ μm | Measured |
| Attachment rate of ParA to nucleoid | $50\ \text{sec}^{-1}$<br>$0.03\ \text{sec}^{-1}$ | (Ietswaart et al., 2014)<br>(Lim et al., 2014) |
| Diffusion constant of ParA on nucleoid | $(0.01 - 1)\ \mu m^2 \text{sec}^{-1}$<br>$0.01\ \mu m^2 \text{sec}^{-1}$ | (Ietswaart et al., 2014)<br>(Lim et al., 2014) |
| ATP-hydrolysis rate of PomZ in contact with PomXY and DNA | $65\ \text{ATP/h} \times \frac{1}{2} \approx 0.01\ \text{sec}^{-1}$ | Measured |
| Diffusion constant ParB/*parS* complex in cytosol | $0.0003\ \mu m^2 \text{sec}^{-1}$<br>$0.0001\ \mu m^2 \text{sec}^{-1}$ | (Ietswaart et al., 2014)<br>(Lim et al., 2014) |
| Spring constant of ParA-ParB bond | $5 \times 10^{-2}\ \text{pN/nm}$<br>$\approx 10^4\ k_B T/\mu m^2$ | (Hu et al., 2015) |
| Generation time | $5h = 18000\ \text{sec}$ | (Treuner-Lange et al., 2013) |
| Time the cluster needs to move to midcell | $(82 \pm 51)\ \text{min (PomY)}$<br>$(81 \pm 53)\ \text{min (PomX)}$ | Measured |

**Table S2**. Parameters used in the simulations

| Parameter | Symbol | Value | Tested parameter space where model functions |
|---|---|---|---|
| Number of PomZ dimers | $n_{total}$ | 100 | |
| Length of nucleoid | L | 5 μm | |
| Diameter of PomXY cluster | $L_{cluster}$ | 0.7 μm | |
| Maximal density of PomZ binding sites on PomXY cluster | $c_{cluster}$ | 1/0.007 μm$^{-1}$ | |
| Maximal density of PomZ binding sites on nucleoid | $c_{nucleoid}$ | 1/0.007 μm$^{-1}$ | |
| Lattice spacing | a | 0.007 μm | |
| Attachment rate of PomZ to nucleoid | $k_{on}$ | 0.1 s$^{-1}$ | $(0.01 - 10)$ s$^{-1}$ |
| Diffusion constant of PomZ on nucleoid | D | 0.01 μm$^2$sec$^{-1}$ | |
| ATP-hydrolysis rate of PomZ bound to nucleoid and cluster | $k_h$ | 0.01 sec$^{-1}$ | |
| Diffusion constant PomXY cluster in cytosol → $\gamma = k_B T/D_{cluster}$ | $D_{cluster}$ | 0.0002 μm$^2$sec$^{-1}$ | |
| Spring constant of PomZ dimer | k | $10^4$ $k_B T/$ μm$^2$ | $(10^3 - 5 \times 10^4)$ $k_B T$ / μm$^2$ |
| Attachment rate of PomZ dimer to cluster in unstretched state | $k_a^0$ | 5 sec$^{-1}$ | $(5 - 500)$ sec$^{-1}$ |
| Initial position of PomXY cluster | p | 0 μm or 0.5 μm | |

**Table S3**. *M. xanthus* strains used in this work

| Strain | Relevant characteristics[1] | Reference |
|---|---|---|
| DK1622 | Wild-type (WT) | (Kaiser, 1979) |
| SA3108 | Δ*pomZ* | (Treuner-Lange et al., 2013) |
| SA3131 | Δ*pomZ*/*attB*::P$_{nat}$ *pomZ-mCherry* (pKA28) | (Treuner-Lange et al., 2013) |
| SA3139 | *ftsZ*+/*attB*::P$_{nat}$ *ftsZ-mCherry* (pKA32), | (Treuner-Lange et al., 2013) |
| SA3146 | Δ*pomZ*/*attB*::P$_{nat}$ *pomZ*$^{D90A}$-*mCherry* (pKA43) | (Treuner-Lange et al., 2013) |
| SA3147 | Δ*pomZ*/*attB*::P$_{pilA}$ *pomZ-mCherry* (pKA45) | This study |
| SA4169 | *pilQ1*, Δ*ftsZ*/P$_{cuoA}$::P$_{cuoA}$ *ftsZ*, *ftsK*+/*attB*::P$_{pilA}$*ftsK-mCherry* (pNG10A*ftsZ*, pDoB12) | This study |
| SA4202 | *parB*+/*attB*:: P$_{nat}$ *parB-eyfp* (pAH7) | (Treuner-Lange et al., 2013) |
| SA4219 | Δ*pomX*, *parB*+/*attB*::P$_{nat}$ *parB-eyfp* (pAH7) | This study |
| SA4223 | Δ*pomX* | This study |
| SA4228 | Δ*pomX*, *ftsZ*+/*attB*::P$_{nat}$ *ftsZ-mCherry* (pKA32) | This study |
| SA4229 | Δ*pomX*/*attB*::P$_{nat}$ *mCherry-PomX* (pKA46) | This study |
| SA4232 | Δ*pomX*, *pomZ*+/*attB*::P$_{nat}$ *pomZ-mCherry* (pKA28) | This study |
| SA4252 | Δ*pomX*/*attB*::P$_{pomZ}$ *mCherry-pomX* (pAH53) | This study |
| SA4254 | Δ*pomXZ* | This study |
| SA4295 | Δ*pomX*/P$_{cuoA}$::P$_{cuoA}$ *mCherry-pomX*, *ftsZ*+/*attB*::P$_{nat}$ *ftsZ-gfp* (pAH52/ pKA51), | This study |
| SA4703 | Δ*pomY* | This study |
| SA4706 | Δ*pomY*, *pomZ*+/*attB*::P$_{nat}$ *pomZ-mCherry* (pKA28) | This study |
| SA4707 | Δ*pomY*, *ftsZ*+/*attB*::P$_{nat}$ *ftsZ-mCherry* (pKA32) | This study |
| SA4709 | Δ*pomY*, *parB*+/*attB*::P$_{nat}$ *parB-eyfp* (pAH7) | This study |
| SA4712 | Δ*pomY*/*attB*::P$_{pilA}$ *pomY-mCherry* (pDS7) | This study |
| SA4713 | Δ*pomY*/*attB*::P$_{nat}$ *pomY-mCherry* (pDS8) | This study |
| SA4718 | *pilQ1*, Δ*ftsZ*/P$_{cuoA}$::P$_{cuoA}$ *ftsZ*, *pomY*+/*attB*::P$_{pilA}$ *pomY-mCherry* (pNG10A*ftsZ*, pDS7) | This study |
| SA4720 | Δ*pomZ*, *pomY*+/*attB*::P$_{pilA}$ *pomY-mCherry* (pDS7) | This study |
| SA4722 | Δ*pomXYZ* | This study |
| SA4734 | Δ*pomY*/P$_{cuoA}$::P$_{cuoA}$ *pomY-mCherry* (pDS21) | This study |
| SA4736 | Δ*pomY*/P$_{cuoA}$::P$_{cuoA}$ *pomY-mCherry*, *ftsZ*+/*attB*::P$_{nat}$ *ftsZ-gfp* (pDS21, pKA51) | This study |
| SA4737 | Δ*pomX*, *pomY*+/*attB*::P$_{pilA}$ *pomY-mCherry* (pDS7) | This study |
| SA4739 | Δ*pomY*, *pomX*+/*attB*::P$_{pomZ}$ *mCherry-pomX* (pAH53) | This study |
| SA4743 | Δ*pomYZ* | This study |
| SA4746 | Δ*mglA*, *pomY*+/*attB*::P$_{pilA}$ *pomY-mCherry* (pDS7) | This study |
| SA4758 | Δ*pomZ*/*attB*::P$_{nat}$ *pomZ*$^{D90A}$-*mCherry*, | This study |
| | pomY+/P*cuoA*::P*cuoA* pomY-eyfp (pKA43, pDS22) | |
| SA4796 | ΔmglA, ΔpomZ, pomY+/attB::P*pilA* pomY-mCherry (pDS7) | This study |
| SA4797 | ΔmglA, ΔpomX/attB::P*pomZ* mCherry-pomX (pAH53) | This study |
| SA4799 | ΔmglA, ΔpomZ/attB::P*pilA* pomZ$^{D90A}$-mCherry (pEB13) | This study |
| SA5000 | ΔpomZ/attB::P*nat* pomZ$^{G62V}$-mCherry (pEB16) | This study |
| SA5001 | ΔpomZ/attB::P*nat* pomZ$^{K66Q}$-mCherry (pKA55) | This study |
| SA5006 | ΔpomZ/attB::P*pilA* pomZ$^{D90A}$-mCherry (pEB13) | This study |
| SA5809 | pilQ1, ΔftsZ/P*cuoA*::P*cuoA* ftsZ, pomX+/attB::P*pomZ* mCherry-pomX (pNG10AftsZ, pAH53) | This study |
| SA5821 | ΔpomZ, pomX+/attB::P*pomZ* mCherry-pomX (pAH53) | This study |
| SA5837 | ΔpomZ/attB::P*nat* pomZ$^{K268E}$-mCherry (pAH100) | This study |
| SA5839 | ΔpomZ/attB:: P*pomZ* mCherry-pomX, P*couA*::P*couA* pomY-eyfp (pAH96, pDS22) | This study |
| SA6130 | ΔpomYX | This study |
| SA6757 | ΔmglA, mxan_4000::tetO-array, mxan18-19::P*van* tetR-eyfp (pMAT13, pMAT112) | This study |
| SA7000 | ΔmglA, ΔpomY/attB::P*pilA* pomY-mCherry (pDS7) | This study |
| SA7008 | ΔmglA, ΔpomY, pomX+/attB::P*pomZ* mCherry-pomX (pAH53) | This study |
| SA7009 | ΔmglA, ΔpomZ, pomX+/attB::P*pomZ* mCherry-pomX (pAH53) | This study |
| SA7011 | ΔmglA, ΔpomZ/attB::P*pilA* pomZ-mCherry (pKA45) | This study |
| SA7014 | ΔpomX, ΔpomZ/attB::P*nat* pomZ$^{D90A}$-mCherry (pKA43) | This study |
| SA7020 | ΔpomZ/attB::P*nat* pomZ-mCherry, pomY+/P*couA*::P*couA* pomY-eyfp (pKA28, pDS22), | This study |
| SA7022 | ΔmglA, ΔpomZ/mxan18-19::P*pilA* pomZ, pomX+/attB::P*pomZ* mCherry-pomX (pDS75, pAH53) | This study |
| SA7027 | ΔmglA, ΔpomZ/mxan18-19::P*nat* pomZ$^{D90A}$, pomY+/attB::P*pilA* pomY-mCherry (pDS80, pDS7) | This study |
| SA7041 | ΔpomX/attB::P*pomZ* mCherry-pomX, pomY+/P*cuoA*::P*cuoA* pomY-eyfp (pAH96, pDS22) | This study |
| SA7042 | ΔpomX/attB::P*pomZ* mCherry-pomX, ssB+/P*nat* ssB::P*nat* ssB-eyfp (pAH96, pAH83) | This study |
| SA7043 | ΔpomY/attB::P*pilA* pomY-mCherry, ssB+/P*nat* ssB::P*nat* ssB-eyfp (pDS19, pAH83) | This study |

[1] Plasmids in brackets contain indicated genes or gene fusions for integration at indicated sites on the genome. Fusions at the *attB* site and the *mxan18-19* intergenic region were expressed from the *pilA* promoter (P$_{pilA}$), their native promoter (P$_{nat}$) or the native promoter of *pomZ* (P$_{pomZ}$). Plasmids containing the *cuoA* promoter (P$_{cuoA}$) were under copper regulated gene expression and were integrated into the P$_{cuoA}$ locus.

**Table S4**. Plasmids used in this work

| Plasmid | Relevant characteristics | Reference/source |
|---|---|---|
| pDS1 | Construct for in-frame deletion of *pomY* | This study |
| pDS3 | Overproduction of PomY-His$_6$ | This study |
| pDS7 | P$_{pilA}$ *pomY-mCherry*, Mx8 *attB*, Km$^R$ | This study |
| pDS8 | P$_{nat}$ *pomY-mCherry*, Mx8 *attB*, Tc$^R$ | This study |
| pDS12 | Construct for in-frame deletion of *pomXYZ* | This study |
| pDS16 | Construct for in-frame deletion of *pomYZ* | This study |
| pDS18 | P$_{nat}$ *pomY-eyfp*, Mx8 *attB*, Tc$^R$ | This study |
| pDS19 | P$_{pilA}$ *pomY-mCherry*, Mx8 *attB*, Tc$^R$ | This study |
| pDS21 | P$_{cuoA}$ *pomY-mCherry*, copper-dependent expression of *pomY-mCherry*, *cuoA*, Km$^R$ | This study |
| pDS22 | P$_{cuoA}$ *pomY-eyfp*, copper-dependent expression of *pomY-eyfp*, *cuoA*, Km$^R$ | This study |
| pDS37 | IPTG-dependent *mCherry-pomX* expression in *E. coli*, Km$^R$ | This study |
| pDS38 | Arabinose-dependent *pomX* expression in *E. coli*, Amp$^R$ | This study |
| pDS43 | IPTG-dependent *pomZ-mCherry* expression in *E. coli*, Km$^R$ | This study |
| pDS45 | Arabinose-dependent *pomY-eyfp* expression in *E. coli*, Amp$^R$ | This study |
| pDS46 | IPTG-dependent *pomY-eyfp* expression in *E. coli*, Km$^R$ | This study |
| pDS68 | IPTG-dependent *pomY-eyfp* and *pomZ-mCherry* expression in *E. coli*, Km$^R$ | This study |
| pDS74 | P$_{pilA}$ *pomZ*, Mx8 *attB*, Tet$^R$ | This study |
| pDS75 | P$_{pilA}$ *pomZ*, *MXAN_18-19* intergenic region, Tc$^R$ | This study |
| pDS80 | P$_{nat}$ *pomZ*$^{D90A}$, *MXAN_18-19* intergenic region, Tc$^R$ | This study |
| pDS81 | pGAD424, *pomY* (C-terminally fused to GAL4-AD), Amp$^R$ | This study |
| pDS82 | pGBT9, *pomY* (C-terminally fused to GAL4-DNA-BD), Amp$^R$ | This study |
| pDS83 | pGAD424, *pomZ* (C-terminally fused to GAL4-AD), Amp$^R$ | This study |
| pDS84 | pGBT9, *pomZ* (C-terminally fused to GAL4-DNA-BD), Amp$^R$ | This study |
| pDS85 | pGAD424, *pomX* (C-terminally fused to GAL4-AD), Amp$^R$ | This study |
| pDS86 | pGBT9, *pomX* (C-terminally fused to GAL4-DNA-BD), Amp$^R$ | This study |
| pDS87 | pGAD424, *ftsZ* (C-terminally fused to GAL4-AD), | This study |

| | Amp^R | |
|---|---|---|
| pDS88 | pGBT9, *ftsZ* (C-terminally fused to GAL4-DNA-BD), Amp^R | This study |
| pAH7 | P_nat *parB-eyfp* Mx8 *attB*, Tc^R | (Treuner-Lange et al., 2013) |
| pAH27 | Construct for in-frame deletion of *pomX*, Km^R | This study |
| pAH35 | P_pilA *mCherry-pomX*, Mx8 *attB*, Km^R | This study |
| pAH52 | P_cuoA *mCherry-pomX*, copper-dependent expression of *mCherry-pomX*, *cuoA*, Km^R | This study |
| pAH53 | P_pomZ *mCherry-pomX*, Mx8 *attB*, Km^R | This study |
| pAH83 | P_nat *ssB-eyfp*, native site, Km^R | This study |
| pAH96 | P_natpomZ *mCherry-pomX*, Mx8 *attB*, Tc^R | This study |
| pAH100 | P_nat *pomZ*^K268E*-mCherry*, Mx8 *attB*, Tc^R | This study |
| pEB13 | P_pilA *pomZ*^D90A*-mCherry*, Mx8 *attB*, Km^R | This study |
| pEB16 | P_nat *pomZ*^G62V*-mCherry*, Mx8 *attB*, Tc^R | This study |
| pSL16 | Construct for in-frame deletion of *mglA*, Km^R | (Miertzschke et al., 2011) |
| pMAT6 | P_cuoA *tetR-eyfp*, Mx8 *attB*, Tc^R | (Harms et al., 2013) |
| pMAT12 | Construct for in-frame deletion of *pomXZ*, Km^R | This study |
| pMAT13 | *tetO*-array+fragment of *mxan4000* region (192°), Km^R | (Harms et al., 2013) |
| pMR3691 | Plasmid for vanillate inducible gene expression | (Iniesta et al., 2012) |
| pMAT56 | pMR3691 with altered multiple cloning site | This study |
| pMAT76 | P_van *tetR-eyfp*, Mx8 *attB*, Tc^R | This study |
| pMAT112 | P_van *tetR-eyfp*, Mx8 *attB*, Str^R | This study |
| pKA1 | Construct for in-frame deletion of *pomZ*, Km^R | (Treuner-Lange et al., 2013) |
| pKA3 | Overproduction of His_6-PomZ, Km^R | (Treuner-Lange et al., 2013) |
| pKA19 | P_pilA *pomZ*, Mx8 *attB*, Km^R | (Treuner-Lange et al., 2013) |
| pKA28 | P_nat *pomZ-mCherry*, Mx8 *attB*, Tc^R | (Treuner-Lange et al., 2013) |
| pKA32 | P_nat *ftsZ-mCherry*, Mx8 *attB*, Tc^R | (Treuner-Lange et al., 2013) |
| pKA43 | P_nat *pomZ*^D90A*-mCherry*, Mx8 *attB*, Tc^R | (Treuner-Lange et al., 2013) |
| pKA45 | P_pilA *pomZ-mCherry*, Mx8 *attB*, Km^R | This study |
| pKA46 | P_nat *mCherry-pomX*, Mx8 *attB*, Tc^R | This study |
| pKA51 | P_nat *ftsZ-gfp*, Mx8 *attB*, Tc^R | (Treuner-Lange et al., 2013) |
| pKA55 | P_nat *pomZ*^K66Q*-mCherry*, Mx8 *attB*, Tc^R | This study |
| pKA70 | Overexpression of FtsZ, Amp^R | (Treuner-Lange et al., 2013) |
| pEMR3 | Overexpression of PomX-His_6, Km^R | This study |
| pGAD424 | P_ADH1 expressed GAL4 activation domain (GAL4-AD), Amp^R | Clontech |
| pGBT9 | P_ADH1 expressed GAL4 DNA binding domain (GAL4-DNA-BD), Amp^R | Clontech |
| pDoB12 | P_pilA *ftsK-mCherry*, Mx8 *attB*, Km^R | This study |

**Table S5.** Primers used in this study

| Primer name | Sequence 5'-3' [1] |
|---|---|
| KA-371 | CCGGAATTCCGACGAGCAGTTGAGCACCAG |
| Mxan_0634-5 | GCGTCTAGAGAGCAGCGCGTCCGGACGCTC |
| KA-373 | GCGTCTAGAGTCACCCCAAGCCATTCC |
| KA-374 | GCGAAGCTTCCTTGAAGTTCAGGAAGAGC |
| KA-224 | GCGGGATCCTCCACGTCCTACGCCTGG |
| Mxan_0636-1 | GCGTCTAGAGGACACGTTCTGTTCAAAGGC |
| Mxan_0636-2 | GCGTCTAGACACCGGGCCCTGAGTCAGGCC |
| Mxan_0636-3 | GCCAAGCTTCAACGCGATGTCCGCGCCC |
| KA-200 | CCGGAATTCCCTGCGCCAACTCCTCAG |
| Mxan_0635-3 | GCGTCTAGAAATCTGCTTCGGGGACACGCC |
| Mxan_0635-1 | GCGTCTAGACAGACGAAGGGCACCCAGCAG |
| Mxan_0635-2 | GCCAAGCTTCCGCTGCACCTCCACCTCGG |
| Mxan_0634-11 | GCGGAATTCGAGCGACGAGCGTC |
| Mxan_0634-12 | CGCAAGCTTAGCGGCGAAGTATTTG |
| Mxan_0634-4 | GGAAGATCTGGCGGAAGCGGCGAAGTATTTGTGCC |
| Mxan_0634-6 | GCGTCTAGAGTGAGCGACGAGCGTCCG |
| mCherry stop rev HindIII | GGGAAGCTTTTACTTGTACAGCTCGTC |
| KA-396 | GCGGGATCCGCCGCCGGCTCCGGCATGGTGAGCAAGGGCGAGG |
| KA-397 | GCCAAGCTTTCACTTGTACAGCTCGTCCATGCC |
| KA-413 | TCCTGAACTTCAAGGTTGGCACCGGCAAGAC |
| KA-414 | GTCTTGCCGGTGCCAACCTTGAAGTTCAGGA |
| KA-302 | CGGGGTACCATGGTGAGCAAGGGCGAGGAG |
| KA-303 | GCGGGATCCGGCGGAGCCCTTGTACAGCTCGTCCATGCC |
| KA-382 | CCGGAATTCGCTCGCGGACTTCCTGTCC |
| KA-383 | CGGGGTACCGGTTCTCAGCCGGCCTGC |
| KA-417 | GGTGGCACCGGCCAGACGTCGCTGT |
| KA-418 | ACAGCGACGTCTGGCCGGTGCCACC |
| AH-91 | ATCCGGCAGTGCACCGAGTTCGCGCAGGCCTCC |
| AH-92 | GGAGGCCTGCGCGAACTCGGTGCACTGCCGGAT |
| KA-288 | GCGTCTAGAATGGAAGCGCCGACGTAC |
| KA-293 | GCCAAGCTTTTACTTGTACAGCTCGTCCAT |
| EB-7 | CGGGGATCCGCCCGGCTGACGCCAGC |
| EB-8 | CGGAATTCAAGCCGCTGCCGCAGTAGAGAC |
| EB-11 | CGTCTAGAGTGAGCAAGGGCGAGGAG |
| EB-12 | CGCGGATCCCTACTTGTACAGCTCGTCCATGC |
| EB-5 | GCCAAGCTTAAGACGCCGGAGGGCTTC |
| EB-6 | GCTCTAGAAATTCGCCAGAACCAGCAGCGGAGCCAGCCGAAGGGGATGTCGTCGTCG |
| Mxan18-19 fwd BsrDI | GCGATCATTGCGCGCCAGACGATAACAGGC |

| Mxan18-19 rev BlpI | GCGGCTGAGCCCGCGCCGACAACCGCAACC |
|---|---|
| Mxan_0636 BspHI fwd | GCGTCATGAAGAAAGCCTTTGAACAGAACG |
| Mxan_0636-3 HindIII rev stop | GCGAAGCTTTCAGCGCACCGTGGCCTGAC |
| mCherry BspHI fwd | GCGTCATGAGCAAGGGCGAGGAGGATAAC |
| Mxan_0635 BspHI fwd | GCGTCATGATGGAAGCGCCGACGTACAGC |
| KA-478 | GCCAAGCTTTCATTACTTGTACAGCTCGTCCAT |
| Mxan_0634-18 | GCGTCATGAGCGACGAGCGTCCGGAC |
| Mxan_0634-15 | GCGAAGCTTTCACTTGTACAGCTCGTCCAT |
| Mxan_0635 start NdeI | GCGCATATGGAAGCGCCGACGTACAGCTCC |
| mCherry stop PacI | GCCTTAATTAATCACTTGTACAGCTCGTCCATG |
| NdeI-pomX fwd | GGAATTCCATATGAAGAAAGCCTTTGAACAG |
| pomX C-term His rev | GCCAAGCTTGCGCACCGTGGCCTGACTCAGG |
| mCherry fwd XbaI | GCGTCTAGAGTGAGCAAGGGCGAGGAG |
| KA-384 | GCGGGATCCGGCGGAGCCATGAAGAAAGCCTTTGAACAG |
| KA-348 | GCCAAGCTTTCAGCGCACCGTGGCCTG |
| pomZ prom rev XbaI | GCGTCTAGACATGGACGACCCTTCGAAAG |
| Mxan_0634 fwd EcoRI | GCGGAATTCGTGAGCGACGAGCGTCCGGAC |
| Mxan_0634 rev stop BglII | GCGAGATCTTCAAGCGGCGAAGTATTTGTG |
| Mxan_0635 fwd SalI linker | GCGGTCGACCTATGGAAGCGCCGACGTACAGC |
| Mxan_0635 rev stop BglII | GCGAGATCTTCAGCCGGCCTGCTGGGTGCC |
| Mxan_0636 fwd EcoRI | GCGGAATTCATGAAGAAAGCCTTTGAACAG |
| Mxan_0636 rev stop BamHI | GCGGGATCCTCAGCGCACCGTGGCCTGACTC |
| Mxan_5597 fwd EcoRI | GCGGAATTCATGGACCAGTTCGATCAGAAC |
| Mxan_5597 rev stop BamHI | GCGGGATCCTTACGGCAGTTCCGTCTGGCC |
| Mxan_0635 fwd BamHI linker | GCGGGATCCGTATGGAAGCGCCGACGTACAGC |
| Mxan_0635 rev stop PstI | GCGCTGCAGTCAGCCGGCCTGCTGGGTGCC |
| ftsK nostop BamHI rev | GCGGGATCCGGCGGAGCCCATGGCCCCGGCGCCGGG |
| ftsK start XbaI | GCTCTAGAACGGCGAAGAAGGGTCG |
| Cla-Sca linker + | TATCGATGAGTACTGGTAC |
| Cla-Sca linker - | AGCTACTCATGAC |

| TetR-YFP-ClaI | GCGCATCGATGTCTAGATTAGATAAAAGT |
| TetR-YFP-KpnI | GCGCGGTACCAAGCTTTTACTTGTACAGC |
| pIJ778 SacI down | CATGAGCTCAGCCAATCGACTGGCG |
| pIJ778 BglII Pst up | GCGCAGATCTCCTGCAGTTCGAA |
| pMAT76 BglII | GCGCAGATCTGGAACGATAGGGACG |
| pMAT76 SacI | GCGCGAGCTCTCAGGACCGCTGCCGGAGC |

[1] Sequences marked in blue indicate restriction sites. Green sequences indicate bases inserted as linker and red sequences start and stop codons.